\documentclass[11pt]{article}
\usepackage{amsthm,amsmath,latexsym,amssymb,amsfonts,amsbsy}
\usepackage{graphicx,lscape,fancyhdr,array,stmaryrd,euscript,wrapfig}
\usepackage{cite,wrapfig,color,youngtab,ytableau}

\usepackage{hyperref}

\newcolumntype{x}[1]{%
>{\centering\hspace{0pt}}m{#1}}%
\newcolumntype{w}[1]{%
>{\raggedright\hspace{0pt}}m{#1}}%
\newcolumntype{z}[1]{%
>{\raggedleft\hspace{0pt}}m{#1}}%

\newcommand{\be}{\begin{equation}}
\newcommand{\ee}{\end{equation}}

\definecolor{rougef}{rgb}{0.56,0,0}
\definecolor{vertf}{rgb}{0,0.5,0}
\definecolor{bleuf}{rgb}{0,0,0.8}
\definecolor{violetf}{rgb}{0.5,0,0.5}

\begin{document}
\renewcommand{\thefootnote}{\fnsymbol{footnote}}
\begin{flushright}
\vspace{1mm}
\end{flushright}

\vspace{1cm}

\begin{center}
{\textbf{\textsc{OFF-SHELL HODGE DUALITIES IN LINEARISED GRAVITY AND $E_{11}$}}}

\vspace{2cm}

\textsc{Nicolas Boulanger\footnote{Research Associate of the Fund for
Scientific Research-FNRS (Belgium); nicolas.boulanger@umons.ac.be}, 
Paul. P. Cook\footnote{paul.cook@kcl.ac.uk} 
and Dmitry Ponomarev\footnote{dmitri.ponomarev@umons.ac.be}}

\vspace{2cm}

{\em${}^{*,\ddagger}$ Service de M\'ecanique et Gravitation, Universit\'e de Mons -- UMONS\\
20 Place du Parc, 7000 Mons (Belgium)}
\vspace*{.5cm}

{\em${}^\dag$ Mathematics Department, King's College London, Strand, London WC2R 2LS, UK}

\vspace{1cm}

\end{center}

\vspace{0.5cm}
\begin{abstract}
In a spacetime of dimension $n\,$, the dual graviton is characterised 
by a Young diagram with two columns, the first of length $n-3$ and the 
second of length one. In this paper we perform the off-shell dualisation relating the dual 
graviton to the double-dual graviton, displaying the precise off-shell field 
content and gauge invariances. We then show that one can further perform
infinitely many off-shell dualities, reformulating linearised gravity in 
an infinite number of equivalent actions. 
The actions require supplementary mixed-symmetry fields which are contained 
within the generalised Kac-Moody algebra $E_{11}$ and are associated with null and imaginary roots.
\end{abstract}
\newpage


\renewcommand{\thefootnote}{\arabic{footnote}}
\setcounter{footnote}{0}

\section{Introduction}\setcounter{equation}{0}
\label{sec:intro}

The identification of symmetries of the low-energy limits
of M-theory is expected to illuminate the complete definition of M-theory. 
Several proposals have been made that identify 
Kac-Moody algebras within supergravity. It was conjectured by West that the non-linear realisation of the generalised 
Kac-Moody algebra $E_{11}$ is an extension of maximal supergravity relevant to M-theory \cite{West:2001as}. 
Explicit constructions were also discovered that exhibited the hyperbolic Kac-Moody algebra $E_{10}$ as a symmetry of 
the equations of motion in the vicinity of a cosmological singularity \cite{Damour:2000hv,Damour:2002et}. 
That affine and hyperbolic Kac-Moody algebras would appear as hidden symmetries of 
supergravity, when the theory is dimensionally reduced, had been anticipated 
previously \cite{Julia:1980gr, Julia:1997cy}.

The Kac-Moody algebra $E_{11}$ may be decomposed into an infinite set of highest weight tensor 
representations of $\mathfrak{sl}_{11}$ which are graded by the level within the decomposition at which they occur. 
At low levels, $E_{11}$ contains a field which West identified with the dual graviton \cite{West:2001as}. 
In a spacetime of dimension $n=11\,$, the dual graviton is characterised 
by a Young diagram with two columns, the first of length $n-3=8$ 
and the second of length one, which we denote by the symbol $C_{[8,1]}$ corresponding to the components 
$C_{\mu_1\ldots\mu_8,\nu}\equiv C_{\mu[8],\nu}\,$.  
In \cite{West:2001as} a first-order action was proposed which is equivalent to nonlinear 
gravity and features an auxiliary field that, after extremising the action with 
respect to the vielbein and linearising the resulting equation, is identified with the curl of the dual graviton. 
It was later shown in \cite{Boulanger:2003vs} that the action given in \cite{West:2001as}, 
linearised and specifying $n=5\,$, lead to the action proposed by Curtright  
in \cite{Curtright:1980yk}, and for arbitrary $n\geqslant 5$ to the action given by Aulakh et al. in \cite{Aulakh:1986cb}.
In the following we will refer to the action for the dual graviton $S_{\rm Curt.}[C_{[n-3,1]}]$ as 
the Curtright action while the Fierz--Pauli action $S_{\rm FP}[h_{[1,1]}]$ \cite{Fierz:1939ix} 
is equivalent to the Einstein--Hilbert action linearised around a Minkowski background of dimension $n\,$. 

To reiterate, the action proposed in \cite{West:2001as}, when linearised\footnote{In 
order to have a nonlinear action principle equivalent to full gravity and 
containing at the same time the graviton and its dual (and not an extra field that reproduces on-shell 
the curl of the dual graviton in the linearisation), one needs to add extra fields. 
This was done in \cite{Boulanger:2008nd}, following a procedure used in the context of 
gauged supergravity. 
Actually, a set of equations equivalent to the field equations derived in \cite{Boulanger:2008nd} 
had been proposed earlier in \cite{West:2002jj}, 
albeit in a different and non-Lagrangian form, and where the extra field had been introduced
with remarkable insight. The results obtained in \cite{Boulanger:2008nd} (where also the gauge structure 
of the theory was clarified),    
therefore strengthen and give an alternative way of understanding the equations found 
in \cite{West:2002jj}.}, 
was shown \cite{Boulanger:2003vs} to reproduce not only the Fierz--Pauli action upon extremising with respect 
to the auxiliary field, but also the Curtright action after extremising with respect to the 
linearised vielbein and substituting the result inside the parent action, thereby demonstrating that
the Fierz--Pauli and Curtright actions are equivalent to each other.

Here we consider all further dualisations of the graviton and first construct the action of the 
double-dual graviton $D_{\mu[n-3],\nu[n-3]}$, a field which was proposed in \cite{Hull:2000rr,Hull:2000zn,Hull:2001iu}, 
whose parent action also reduces to the Curtright action upon algebraic elimination of one of its fields. 
The procedure will be described for an infinite set of further dualisations of the graviton giving rise to three infinite 
gravity towers with fields $\tilde{h}_{\mu_1[n-2],\ldots,\mu_k[n-2],\nu,\rho}\,$, 
$\tilde{C}_{\mu_1[n-2],\ldots,\mu_k[n-2],\nu[n-3],\rho}$
and $\tilde{D}_{\mu_1[n-2],\ldots,\mu_k[n-2],\nu[n-3],\rho[n-3]}$ ($k=1,2,\ldots$)  
which we refer to as the Fierz--Pauli tower, the dual graviton tower and 
the double-dual tower, respectively. 
We note that the fields entering what we call here the dual graviton tower were 
first recognised as an infinite set of dual gravitons in \cite{Riccioni:2006az}.
{}From the work in the present paper the reader 
will be able to construct the actions for the fields in any of the 
infinite towers.

It has been proposed by Hull \cite{Hull:2000rr,Hull:2000zn,Hull:2001iu} that the further off-shell dualisation of the 
Fierz--Pauli graviton, if it is possible, should unveil some hidden symmetries of M-theory that had gone unnoticed before.
Hull conjectured a duality between an exotic six-dimensional $(4,0)$ 
superconformal theory and the strong coupling limit of maximally
supersymmetric ${\cal N}=8$ supergravity in 5 dimensions. 
Upon dimensional reduction of the field content of the linearised 
six-dimensional theory down to five dimensions~\cite{Hull:2000rr,Hull:2000zn,Hull:2001iu}, 
not only does the dual graviton $\tiny\yng(2,1)$ appear 
but also a double-dual graviton $\tiny\yng(2,2)\,$. 
The exotic interacting six-dimensional theory suggested by Hull is to maximally  
${\cal N}_5=8$ supergravity what the superconformal ${\cal N}_6=(2,0)$ theory 
is to maximally supersymmetric Yang--Mills theory in five dimensions. 
It is very tempting to think, like Hull, that there is a corner of M-theory 
that contains the exotic ${\cal N}_6=(4,0)$ theory. 
Note that this theory has been discussed recently in 
\cite{Chiodaroli:2011pp}.

In the present paper we show that it is indeed possible to nontrivially 
dualise Curtright's action,
thereby making contact with Hull's double-dual graviton. 
The off-shell and manifestly covariant formulation of the double-dual
graviton that we obtain is less economical than Curtright's dual 
formulation or the Fierz--Pauli original one, in the sense that the 
off-shell spectrum is larger, although on-shell the degrees of freedom
are the same by construction.
We show that it is actually possible to perform infinitely many further
off-shell Hodge dualisations to obtain what we call here 
the Fierz--Pauli, dual and double-dual graviton towers, 
thereby describing linearised gravity in less and less economical 
formulations. 
These descriptions, on the other hand, allow us to make explicit contact 
with generators of $E_{11}\,$. 

In \cite{Riccioni:2006az} fields of exactly the same symmetry types as those entering
the dual graviton tower, e.g. containing a Young tableau of the type $\tilde{C}_{\mu_1[n-2],\ldots,\mu_k[n-2],\nu[n-3],\rho}$ 
for $n=11\,$, were identified within the generalised Kac-Moody algebra $E_{11}\,$. 
It was conjectured therein that this infinite tower of $\mathfrak{sl}_{11}$ representations 
contained all possible on-shell dual descriptions of the graviton. 
The work in this paper supports the interpretation that 
the tower of $\mathfrak{sl}_{11}$ representations identified in
\cite{Riccioni:2006az}
indeed contains dual gravitational fields 
and places it on a firm off-shell footing. 
Additionally we identify further infinite gravity towers 
which are not contained in $E_{11}$. We demonstrate how these dual fields may be incorporated at the level of an action via 
Hodge duality. The actions require sets of supplementary fields, described by mixed symmetry tensors, which are all 
contained in $E_{11}$ and associated with null or imaginary roots. 

The fields in what we call here the dual graviton tower 
have also been 
interpreted in \cite{Englert:2007qb} as exotic gravitational solutions where it was shown that each of these exotic gravity 
fields are related to each other by the Geroch group \cite{Geroch:1970nt,Geroch:1972yt,Breitenlohner:1986um}. Additionally 
the dual graviton tower of fields has been argued \cite{Damour:2002cu} in the context of $E_{10}$ to be related to spatial 
derivatives of the field $C_{\mu[n-3],\nu}$.

In Section \ref{sec:Lagrangian} we explicitly construct a parent action 
that on the one hand reduces to Curtright's action upon eliminating some 
auxiliary fields, and on the other hand produces a new action that 
features Hull's double-dual graviton upon
eliminating another set of auxiliary fields.
We analyse the gauge invariance of the aforementioned new action. 

In Section \ref{sec:Mtimes} we show how infinitely many dual formulations
of Fierz--Pauli theory can be obtained, which become less and less economical in the
sense that the off-shell spectrum entering the successive Lagrangians 
gets bigger and bigger.

In Section \ref{sec:E11 notes} we relate the field content of the 
double-dual action of Section \ref{sec:Lagrangian} containing the spectrum of 
the $(4,0)$ linearised theory proposed by Hull with the generators of $E_{11}\,$.
The gravitational degrees of freedom present in the five-dimensional theory are traced to their six-dimensional origin and 
the double-dual graviton of Hull is shown not to be present within $E_{11}\,$. 
The gravitational degrees of freedom are instead identified with an exotic gravity field within the dual tower of fields 
of $E_{11}$ associated with the dual graviton. Furthermore the off-shell fields required for the action constructed in 
Section \ref{sec:Mtimes} are identified with the null and imaginary roots of $E_{11}$ and appear at the same level in the 
algebraic decomposition as the exotic dual gravity field in question.

We conclude the paper with a summary of our work and some comments in Section \ref{sec:Conclusion}.

\section{On-shell and off-shell double dualisation} 
\label{sec:Lagrangian}

\paragraph{Preamble: on-shell dualisation.}

Before attacking the problem of off-shell dualisation of linearised 
gravity beyond the Curtright level, where the dual graviton appears, we first show that it is very simple 
to write a quadratic action that features both the double-dual field 
$D_{\mu[n-3],\nu[n-3]}$ and the Fierz--Pauli field $h_{\mu\nu}$ and 
produces, on-shell, the appropriate duality relation between them 
together with the Fierz--Pauli equations. Such a dualisation is however
not satisfactory since it takes place only on-shell 
instead of off-shell, nevertheless it prepares the grounds for the 
rest of the paper and allows us to introduce our notation. 
Taking $n=5$ for the sake of clarity, one considers the following action\footnote{We
choose the space-time signature to be $(-+\cdots +)\,$.
The epsilon symbol is defined by $\varepsilon^{012\cdots}=+1\,$. 
We denote strength-one antisymmetrisation of indices by square 
brackets and the components of an antisymmetric tensor 
$T_{a_1\ldots a_p}=T_{[a_1\ldots a_p]}$ will sometimes be denoted 
by $T_{a_1 \ldots a_p}= T_{a[p]}\,$.
Similarly, we sometimes use 
$\partial_{\mu} e_{\mu,\nu} \equiv \frac{1}{2}\, 
(\partial_{\mu_1} 
e_{{\mu_2},\nu} -\partial_{\mu_2} e_{\mu_1,\nu})\,$. 
Differential form degree is denoted by a subscript in boldface font, 
so that $e_{\mathbf{2}}^{a[2]}\,$ denotes a two-form taking its values
in the antisymmetric rank-two irreducible representation of the Lorentz algebra.
The (flat) background vielbein is denoted by $\bar{h}_{\mathbf{1}}^a\,$.}
\begin{align}
 S[h_{\mu\nu},D_{\mu\nu,\rho\sigma}] = \int d^5x \; 
\varepsilon^{\mu\nu\rho\sigma\lambda}\varepsilon_{\alpha\beta\gamma\delta\epsilon}
\;D_{\mu\nu,}{}^{\alpha\beta}\,\left( R_{\rho\sigma,}{}^{\gamma\delta}(h)\,
\delta^{\epsilon}_{\lambda} -\tfrac{1}{2}\,\partial_{\rho}\partial^{\gamma}
D^{\delta\epsilon,}{}_{\sigma\lambda} \right)
\label{actionphiD}
\end{align}
where 
$R_{\mu\nu}{}^{\rho\sigma}(h) = 2\; \partial^{[\rho}\partial_{[\mu}h_{\nu]}{}^{\sigma]}$
is the linearised Riemann tensor of the field $h_{\mu\nu}\,$. 
The gauge invariances read
\begin{align}
 \delta h_{\mu\nu} &= 2\,\partial_{(\mu}\epsilon_{\nu)}\quad, 
 & \delta D_{\mu\nu,\rho\sigma} &= 
 (\partial_{\mu}\lambda_{\rho\sigma,\nu} - \partial_{\nu}\lambda_{\rho\sigma,\mu}) +
 (\partial_{\rho}\lambda_{\mu\nu,\sigma} -  \partial_{\sigma}\lambda_{\mu\nu,\rho})
\end{align}
where $\lambda_{\mu\nu,\rho}$ is an irreducible $\mathfrak{gl}_5$ tensor of type $[2,1]\,$, 
\emph{i.e.} it obeys
$\lambda_{\mu\nu,\rho}=-\lambda_{\nu\mu,\rho}\,$, $\lambda_{[\mu\nu,\rho]}=0\,$. 
As for the field equations, introducing the linearised curvature for the double dual graviton 
as $K_{\mu\nu\rho,}{}^{\alpha\beta\gamma} = 
 12 \,\partial^{[\gamma}\partial_{[\rho}D_{\mu\nu],}{}^{\alpha\beta]}\,$, we have
\begin{align}
&\left\{ 
\begin{array}{c}
\frac{\delta S[h,D]}{\delta D^{\mu\nu,\rho\sigma}}\; = \;0   \\
\frac{\delta S[h,D]}{\delta h^{\mu\nu}}\; = \;0    
\end{array}\right.
& \Longleftrightarrow &
&\left\{ 
\begin{array}{c}
 R_{\mu\nu,\alpha\beta}(h)  = 
  \varepsilon_{\mu\nu\rho\sigma\lambda} \;K^{\rho\sigma\lambda,\gamma\delta\epsilon} \;
   \varepsilon_{\alpha\beta\gamma\delta\epsilon} \\
  \eta^{\rho\gamma} \eta^{\sigma\delta}  \;K_{\mu\rho\sigma,\nu\gamma\delta} =  \;0    
\end{array}\right.\quad , 
\end{align}
which can be written in the hyperform notation \cite{Bekaert:2002dt} as 
$R_{[2,2]} = *_1 *_2 K_{[3,3]}\,$ and $ {\rm{Tr}}^2 K_{[3,3]} = \;0\;$,  
or equivalently as
  $R_{[2,2]} = *_1 *_2 K_{[3,3]}\,$ and $ {\rm{Tr}} R_{[2,2]} = \;0\,$
where the last equation is the Fierz--Pauli equation for the spin-2 field $h_{\mu\nu}\,$. 
The above equations were discussed by Hull in \cite{Hull:2000rr,Hull:2000zn,Hull:2001iu}.  
\vspace*{.5cm}

\noindent
The action $S[h, D]$ (\ref{actionphiD}) can be written in the frame-like formalism of 
\cite{Zinoviev:2003ix,Skvortsov:2008vs,Skvortsov:2008sh,Skvortsov:2010nh}. 
Since the first-order action for the $D_{[2,2]}$-field in five dimensions is
(see e.g. Section 3 of \cite{Zinoviev:2003ix})
\begin{align}
 S[e_{\mathbf{2}}^{a[2]},\omega_{\mathbf{2}}^{a[3]}] = \int_{{\cal M}_5} 
\epsilon_{a[5]} \left( \tfrac{3}{8}\;\bar{h}_{\mathbf{1}}^a \wedge\omega_{\mathbf{2}}^{a[2]c}\wedge 
\omega_{\mathbf{2}}^{a[2]}{}_c + 
\tfrac{1}{12}\; e_{\mathbf{2}}^{a[2]}\wedge {\rm d}\omega_{\mathbf{2}}^{a[3]}   \right) \quad,
\end{align}
we can directly write down the action (\ref{actionphiD}) in the following hybrid form:
\begin{align}
\label{hybrid}
 S[h_{\mathbf{1}}^{a},e_{\mathbf{2}}^{a[2]},\omega_{\mathbf{2}}^{a[3]}] = \int_{{\cal M}_5} 
\epsilon_{a[5]} \left( \tfrac{3}{8}\;\bar{h}_{\mathbf{1}}^a \wedge\omega_{\mathbf{2}}^{a[2]c}\wedge 
\omega_{\mathbf{2}}^{a[2]}{}_c + 
\tfrac{1}{12}\;   e_{\mathbf{2}}^{a[2]}\wedge
{\rm d} \left[\omega_{\mathbf{2}}^{a[3]} + \bar{h}_{\mathbf{1}}^a \,\omega_{\mathbf{1}}^{a[2]}
(h_{\mathbf{1}})\right]   \right) 
\end{align}
where $\omega_{\mathbf{1}}^{a[2]}(h_{\mathbf{1}})=dx^{\mu}\omega_{\mu}^{a[2]}(h_{\mathbf{1}})$ 
is the spin connection one-form viewed as a function of the dynamical vielbein fluctuation 
$h_{\mathbf{1}}^a = dx^{\mu}h^a_\mu$ via 
the solution of the linearised zero-torsion
${\rm d} h_{\mathbf{1}}^{a} + {\omega_{\mathbf{1}}}^{a}{}_{b}\;\bar{h}_{\mathbf{1}}^b=0\,$ condition.
The action (\ref{hybrid}) is truly first-order in the sector of the $D_{[2,2]}$ field.
In the spin-2 sector, it features the spin-connection as a function as the vielbein fluctuation, 
as in the linearisation of the Einstein--Cartan--Weyl action [the Einstein--Cartan--Weyl action
is recalled in (\ref{EH}) below]. 
The action (\ref{hybrid}) lends itself to non-linear completion where one replaces everywhere 
the background vielbein $\bar{h}$ by 
$e_{\mathbf{1}}^{a} = \bar{h}_{\mathbf{1}}^{a} + h_{\mathbf{1}}^{a}$ and  
${\rm d} \omega_{\mathbf{1}}^{a[2]}(h_{\mathbf{1}})$ 
by the full non-linear curvature 
$R_{\mathbf{1}}^{a[2]}(e_{\mathbf{1}}) = 
{\rm d} \omega_{\mathbf{1}}^{a[2]}(e_{\mathbf{1}}) + \omega_{\mathbf{1}}^{ab}(e_{\mathbf{1}})
{\omega_{\mathbf{1}}}_b{}^{a}(e_{\mathbf{1}})\,$.  
\vspace*{.2cm}

Although the action proposed above has the advantage of relating the double-dual 
field introduced by Hull to the usual graviton, the dualisation relation is only 
obtained on-shell which is not sufficient for a genuine equivalence of theories 
and for the purpose of quantisation \cite{Fradkin:1984ai}. 
One needs a parent action that relates the Fierz--Pauli action to a new action
that would incorporate the double-dual field $D_{a[n-3],b[n-3]}\,$. 
We construct the parent action in the sequel. 

\paragraph{Off-shell dualisation: first round.}
We start by reviewing the off-shell dualisation of the graviton as given 
in \cite{West:2001as,Boulanger:2003vs}.
For this one uses the fact that the second-order Einstein-Hilbert action based on the vielbein 
$e_{\mu}{}^{a}$ can be written, up to boundary terms, as \cite{Weyl:1929fm} 
\begin{align}
\label{EH}
  S_{\rm EH}[e_{ab}] \ = \ -\int d^n x\hspace{0.1em}e\left(\Omega^{ab,c}\,\Omega_{ab,c}+
  2\,\Omega^{ab,c}\,\Omega_{ac,b} - 4\, \Omega_{ab,}{}^b\,\Omega^{ac,}{}_{c}\right)\;,  
 \end{align}
where
\begin{align}
 \Omega_{ab,}{}^{c}\ = \ \Omega_{ab,}{}^{c}(e) \ = \  2\,e_{a}{}^{\mu}\,e_{b}{}^{\nu}\,\partial_{[\mu}e_{\nu]}{}^{c}
\end{align}
are the coefficients of anholonomicity. 
This form of the Einstein-Hilbert action can be recast into first-order form by
introducing an auxiliary field $Y_{ab,c}=-Y_{ba,c}\,$,
\begin{align}
\label{first}
  S[Y_{ab,c},e_{ab}] \ = \ -2\int d^nx \; e
  \left(Y^{ab,c}\Omega_{ab,c}(e)-\tfrac{1}{2}\,Y_{ab,c}Y^{ac,b}+\tfrac{1}{2(n-2)}\,Y_{ab,}{}^{b}Y^{ac,}{}_{c}\right)\;. 
\end{align}
The field equation of $Y$ can be used to solve for it in terms of $\Omega(e)\,$,
\begin{align}
\label{Ysol}
  Y_{ab,c}(e) \ = \
  \Omega_{ab,c}-2\Omega_{c[a,b]}+4\eta_{c[a}\Omega_{b]d,}{}^{d}\;.  
\end{align}
After reinserting (\ref{Ysol}) into (\ref{first}), one precisely
recovers the Einstein-Hilbert action in the form (\ref{EH}). In
fact, the action (\ref{first}) coincides with the standard first
order action with the spin connection as independent field, up to a
mere field redefinition which replaces the spin connection by
$Y_{ab|c}\,$. For later use one notes that (\ref{first}) has the same
symmetries as the original Einstein-Hilbert action. First, it is
manifestly diffeomorphism invariant. Moreover, the invariance of the
second-order action (\ref{EH}) under the local Lorentz group can be
elevated to a symmetry of the first-order action by requiring that
the auxiliary field $Y_{ab|c}$ transforms as
\begin{align}
\label{auxsym}
  \delta_{\Lambda}Y_{ab,c} \ = \
  -2\,e_{c}{}^{\mu}\partial_{\mu}\Lambda_{ab}-4\eta_{c[a}e^{\mu
  d}\partial_{\mu}\Lambda_{b]d}-2\Lambda^{d}{}_{[a}Y_{b]d,c}
  +\Lambda^{d}{}_{c}Y_{ab,d}\;. 
\end{align}
In order to obtain the dual graviton from (\ref{first}) one has to
consider the linearised theory and vary with respect to the metric.
Before linearising, it turns out to be convenient to first rewrite
the action in terms of the Hodge dual of $Y^{ab,c}\,$: 
\begin{align}
\label{dualY}
  Y^{ab,c} \ = \ \tfrac{1}{(n-2)!}\,\epsilon^{abc_1\cdots c_{n-2}}
  Y_{c_1\cdots c_{n-2},}{}^{c}\;. 
\end{align}
This yields 
\begin{align}
\label{firstdual}
 \begin{split}
  S[Y,e] \ = \ -\tfrac{2}{(n-2)!}\int
  d^nx\hspace{0.1em}e\Big(&\epsilon^{abc_1\ldots
  c_{n-2}}\,Y_{c_1\ldots c_{n-2},}{}^{c}\,\Omega_{ab,c}+\tfrac{n-3}{2(n-2)}\;
  Y^{c_1\ldots c_{n-2},b}Y_{c_1\ldots c_{n-2},b}\\
  &-\tfrac{n-2}{2}\;Y^{c_1\ldots c_{n-3}a,}{}_{a}\,Y_{c_1\ldots c_{n-3}b,}{}^{b}
  +\tfrac{1}{2}\;Y^{c_1\ldots c_{n-3}a,b}\, Y_{c_1\ldots c_{n-3}b,a}\Big)\;.
 \end{split} 
\end{align}
In the linearisation around flat space,
$e_{\mu}{}^{a}=\delta_{\mu}{}^{a}+\kappa\, h_{\mu}{}^{a}\,$, 
one can ignore the distinction between flat and curved indices. 
In particular, one has now
$\Omega_{\mu\nu,\rho}=2\,\partial_{[\mu}h_{\nu]\rho}\,$, where the
field $h_{\mu\nu}$ still has no symmetry.
The field equation for $h_{\mu\nu}$ is
 \begin{align}
\label{inte}
  \partial_{[\mu_1}Y_{\mu_2\ldots\mu_{n-1}],\nu} \ = \ 0\;.  
\end{align}
The Poincar\'e lemma then implies that $Y$ is the curl of a
potential $C_{\mu_1\ldots\mu_{n-3},\nu}$ (the dual graviton),
that is completely antisymmetric in its first $n-3$ indices but has no definite
$\mathfrak{gl}_n$ symmetry otherwise:
\begin{align}
\label{intesol}
  Y_{\mu_1\ldots\mu_{n-2},\nu} \ = \
  \partial_{[\mu_1}C_{\mu_2\ldots\mu_{n-2}],\nu}\;.  
\end{align}
Inserting this back into the linearisation of (\ref{firstdual}) yields a consistent
quadratic action $S[C]$ for the dual graviton that is equivalent to the Curtright action 
\cite{Curtright:1980yk}.
\vspace*{.5cm}

Up to now, $C_{\mu_1\ldots\mu_{n-3},\nu}$ as defined by
(\ref{intesol}) does not transform in any irreducible ${\mathfrak{gl}}_n$
representation since $Y$ does not possess any irreducible Young-diagram symmetry. 
However, one may check that, after inserting (\ref{intesol})
into the linearisation of (\ref{firstdual}), the resulting action
$S[C]$ is invariant under the following shift symmetry
\begin{align}
\label{stuckel}
  \delta_{\Lambda}C_{\mu_1\ldots\mu_{n-3},\nu} \ = \
  -\Lambda_{\mu_1\ldots\mu_{n-3}\nu}\;,  
\end{align}
with completely antisymmetric shift parameter. Therefore, the
totally antisymmetric part of $C_{\mu_1\ldots\mu_{n-3},\nu}$ can be
gauge-fixed to zero, giving rise to the dual
graviton with a $[n-3,1]$ Young-diagram symmetry. 
In other words, in the action $S[C]$ the dual
graviton appears in a way similar to the graviton in the second-order 
Weyl action (\ref{EH}). One can formulate a genuinely first-order 
action principle for arbitrary $\mathfrak{gl}_n$-irreducible mixed-symmetry 
gauge fields in flat background. 
This has been done by Skvortsov in \cite{Skvortsov:2008sh}. 
The latter formulation is the analogue of the vielbein formalism of gravity 
(\ref{first}) in which the Lorentz transformations act as St\"uckelberg 
transformations and where the spin-connection is viewed as an off-shell independent field. 
\vspace*{.5cm}

Let us mention that, even though (\ref{first}) and thus
(\ref{firstdual}) are first-order formulations of
\textit{non-linear} Einstein gravity, the identification of the dual
graviton in (\ref{intesol}) is only possible in the linearisation,
since in the full theory the integrability condition (\ref{inte}) is
violated \cite{West:2001as}. This is in agreement with the fact
that there is no local, manifestly Lorentz-invariant and non-abelian 
self-interacting theory for the dual graviton~\cite{Bekaert:2002uh,Bekaert:2004dz}.
\vspace*{.3cm}

\paragraph{Towards a second off-shell dualisation.} 

In order to address the problem of a further dualisation, it is useful to introduce the following
quadratic parent action \cite{Boulanger:2003vs}: 
\begin{align}
\label{parent2}
  S[\Omega_{ab,c},Y_{abc,d}] \ = \ - \int d^n x \; \left( 2 \,\Omega_{ab,c}\partial_d Y^{dab,c} 
  + \Omega^{ab,c}\Omega_{ab,c} + 2\,\Omega^{ab,c}\Omega_{ac,b} 
  - 4\,\Omega_{ab,}{}^b\Omega^{ac,}{}_{c}\right)\;. 
 \end{align}
The field $Y_{abc,d}=Y_{[abc],d}$ is a Lagrange multiplier for the constraint 
$\partial_{[d}{\Omega}_{ab],c}=0\,$, implying ${\Omega}_{ab,c}=\partial_{[a}h_{b]c}\,$
where $h_{\mu\nu}$ has no definite symmetry in its two indices. 
Eliminating $Y$ that way, one finds that the action (\ref{parent2}) becomes the 
Einstein--Hilbert action (\ref{EH}) taken at quadratic order, namely the Fierz--Pauli action. 
On the other hand, 
 $\Omega_{ab,c}$ is an auxiliary field and can be eliminated from the action (\ref{parent2})
using its algebraic equations of motion. The resulting action is then the action 
(\ref{firstdual}) at quadratic order, in turn equivalent to the Curtright action.
\vspace*{.5cm}

The action (\ref{parent2}) makes the first dualisation step transparent: 
Knowing the Fierz--Pauli action $S^{FP}[h_{ab}]$ given by the last three terms of 
(\ref{parent2}) with $\Omega=\Omega(h)\,$, one introduces a new field $Y^{abd,c}=Y^{[abd],c}$ 
that, via the first term $2 \int \,\Omega_{ab,c}\partial_d Y^{dab,c}$ enforces the relation 
${\Omega}_{ab,c}=2\,\partial_{[a}h_{b]c}\,$ between the first connection of the spin-2 field 
and the spin-2 field itself in the frame formulation of (linearised) gravity
where the vielbein has no definite symmetry in its indices.
Then, eliminating the connection $\Omega$ from the action, one obtains a new action which 
is by construction equivalent to the Fierz--Pauli action and yields indeed 
the Curtright action in a formulation where the off-shell dual field 
$C_{\mu_1\ldots\mu_{n-3},\nu}$ has no definite $\mathfrak{gl}_n$ symmetry.
One can then check that there is a shift symmetry 
(\ref{stuckel}) that ensures that only the $\mathfrak{gl}_n$-irreducible $[n-3,1]$ part 
contributes to the action.
\vspace*{.5cm}

Accordingly, in order to understand a second dualistaion at the level of the action, 
one has to follow the following procedure:
\begin{itemize}
 \item[(i)] Construct the putative action:
\begin{align}
\label{putative}
S^{put.}[H^{a[n-3],}{}_{bc},D^{bcd},{}_{a[n-3]}]  = \int d^nx\; \left[ H^{a[n-3],}{}_{bc}\;
\partial_{d}D^{bcd},{}_{a[n-3]} + `` H H '' \right ]\;
\end{align}
where the part denoted ``$H\,H$'' should give the Curtright action 
via the substitution 
$H_{\mu[n-3],}{}^{\nu[2]} \;\longrightarrow \;2\,\partial^{[\nu_1}C_{\mu[n-3],}{}^{\nu_2]}\,$. 
In other words, a necessary condition is that the Curtright action admits a formulation 
$S_{\rm Curt.}[H(C)]$ in which the ${\mathfrak{gl}}_n$-reducible field $C_{\mu[n-3],\nu}$ 
appears only thought the quantity 
 $H_{\mu[n-3],}{}^{\nu[2]}(C) := 2\,\partial^{[\nu_1}C_{\mu[n-3],}{}^{\nu_2]}\,$;
\item[(ii)] 
Then, supposing that (i) is possible, the field $D^{b[3]},{}_{a[n-3]}$ can be 
eliminated from the action (\ref{putative}), enforcing the relation 
$H_{\mu[n-3],}{}^{\nu[2]} = 2\,\partial^{[\nu_1}C_{\mu[n-3],}{}^{\nu_2]}\,$
and thereby producing the Curtright action $S_{\rm Curt.}[H(C)]\,$.
Alternatively, one can extremise the action (\ref{putative})
with respect to the auxiliary field $H^{a[n-3],}{}_{b[2]}$ and get an action 
\begin{align}
\label{put2}
S[D^{bcd},{}_{a[n-3]}]  = \int d^nx\; \left[ \partial^{e}D_{bce},{}^{a[n-3]} \;
\partial_{d}D^{bcd},{}_{a[n-3]} + \cdots \right ]
\end{align}
which would, by construction, be equivalent to the Curtright action $S_{\rm Curt.}[H(C)]\,$;
\item[(iii)] 
Decomposing the $D$ field into its irreducible $\mathfrak{gl}_n$ components 
\begin{eqnarray}
&D_{b[3]},{}^{a[n-3]} \;=\; X_{b[3]},{}^{a[n-3]} + Z_{b[3]},{}^{a[n-3]} \quad, &
\\
&Z_{b[3]},{}^{a[n-3]} \; :=\;  
\delta_{[b_1}^{[a_1} Z^{(1)}{}_{b_2b_3]},{}^{a_2\ldots a_{n-3}]} 
+ \delta_{[b_1}^{[a_1}\delta_{b_2}^{a_2} Z^{(2)}{}_{b_3]},{}^{a_3\ldots a_{n-3}]} 
+ \delta_{[b_1}^{[a_1}\delta_{b_2}^{a_2}\delta_{b_3]}^{a_3} Z^{(3)}{}^{a_4\ldots a_{n-3}]}
\quad,& 
\\
&  X_{b_1b_2b_3},{}^{b_1a[n-4]} \;\equiv \; 0 \;\equiv \; Z^{(1)}{}_{b_1b_2},{}^{b_1\,a[n-5]}\;, 
        \quad Z^{(2)}{}_{b},{}^{ba[n-6]} \;\equiv\; 0  \;, &  
\end{eqnarray}
one obtains an action containing Hull's two-column $\mathfrak{gl}_n$-irreducible gauge field 
$$D_{a[n-3],b[n-3]}:=\frac{1}{(n-3)!}\; \epsilon_{c[3]a[n-3]} \;X^{c[3]},{}_{b[n-3]}$$ 
(the `double-dual graviton') provided that the components $Z_{b[3]},{}^{a[n-4]}$
of $D_{b[3]},{}^{a[n-3]}$ disappear from the action (\ref{put2}).
If none of the $Z$ components disappear from the action, one would
get an action equivalent to the Curtright action and expressed in terms of the set of 
$\mathfrak{gl}_n$-irreducible gauge fields 
$$\{ D_{a[n-3],b[n-3]}, E^{(1)}{}_{a[n-2],b[n-4]}, E^{(2)}{}_{a[n-1],b[n-5]}, Z^{(3)}{}_{b[n-6]} \}\,,$$
 where 
\begin{eqnarray}
E^{(1)}{}_{a[n-2],b[n-4]}&:=&\frac{1}{(n-2)!}\; \epsilon_{c[2]a[n-2]} \;Z^{(1)\;c[2]},{}_{b[n-4]}\,,
\\ 
E^{(2)}{}_{a[n-1],b[n-5]}&:=&\frac{1}{(n-1)!}\; \epsilon_{ca[n-1]} \;Z^{(2)\;c},{}_{b[n-5]}\,. 
\end{eqnarray}
\end{itemize}
\vspace*{.3cm}

\noindent We now follow this programme and show that it is indeed possible to perform a second dualisation
of the Fierz--Pauli action, the upshot being that all the following fields 
\begin{eqnarray}
\label{spectrum}
\{ D_{a[n-3],b[n-3]}, E^{(1)}{}_{a[n-2],b[n-4]}, E^{(2)}{}_{a[n-1],b[n-5]}, Z^{(3)}{}_{b[n-6]} \}\,, 
\end{eqnarray}
enter the double-dual formulation.  
We will first achieve this programme in the case $n=5\,$ simply in order 
to simplify the formulae and the presentation and then give the results 
in the general $n$-dimensional case, where $n\geqslant 5\,$.
In the case $n=4\,$, it is obvious that one keeps on 
reproducing Fierz--Pauli action, as was already explained in 
\cite{Boulanger:2003vs}.
\vspace*{.3cm}

The first thing to do according to point (i) is to reformulate Curtright's action in terms of the 
quantity $H_{\mu\nu,}{}^{\rho\sigma}(C)=2\,\partial^{[\rho}C_{\mu\nu,}{}^{\sigma]}\,$. 
Taking into account that fact that the action should contain the kinetic term 
$\int d^5x \; \frac{1}{2}\; H_{\mu\nu,}{}^{\rho\sigma}(C) H^{\mu\nu,}{}_{\rho\sigma}(C)$ and should
be invariant under the transformations
\begin{align}
\delta C_{\mu\nu,\rho} = 2\,\partial_{[\mu}\xi_{\nu]\rho} 
        + \tfrac{1}{2}\,\Lambda_{\mu\nu\rho}\quad ,
\label{gaugetransfoC}
\end{align}
where $\xi_{\nu\rho}$ has no symmetry in its two indices while 
$\Lambda_{\mu\nu\rho}=\Lambda_{[\mu\nu\rho]}\,$, one can write all the possible quadratic 
terms in the action and fix the free coefficients in order to ensure the invariance under 
(\ref{gaugetransfoC}). The procedure is direct and gives the following 
result\footnote{As a consistency check one can take this action, 
integrate by parts, where necessary, and show that it gives back the usual 
Curtright action up to boundary terms. This works indeed as expected.
Notice that not all the possible terms bilinear in $H$ have been used.
The term $H^{\nu_1\nu_2,\rho_1\rho_2}H_{\rho_1\rho_2,\nu_1\nu_2}\,$, is omitted because it is redundant when $H=H(C)$.} 
\begin{eqnarray}
S_{\rm Curt.}[H_{\mu\nu,}{}^{\rho\sigma}(C)]  = \int d^5x &\Big[& \tfrac{1}{2}\, 
H_{\mu\nu,}{}^{\rho\sigma} \,H^{\mu\nu,}{}_{\rho\sigma} + 
H^{\mu\nu,}{}_{\rho\sigma}\,H_{\mu}{}^{\rho,}{}_{\nu}{}^{\sigma}\;
- 3\, H^{\mu\nu,}{}_{\rho\nu}H_{\mu\sigma,}{}^{\rho\sigma}
\nonumber \\
&  & -\;  H^{\mu\nu,}{}_{\rho\nu}H^{\rho\sigma,}{}_{\mu\sigma} 
+ H_{\mu\nu,}{}^{\mu\nu} H_{\rho\sigma,}{}^{\rho\sigma} \Big] 
 \;=\; \int d^5x \;{\cal{L}}(H(C)) \; .\qquad
\label{CurtH}
\end{eqnarray}
Then, complying with point (ii) in the programme above, one considers the following 
parent action
\begin{align}
S[D^{\rho\sigma\lambda,}{}_{\mu\nu},H^{\mu\nu,}{}_{\rho\sigma}] \;=\; 
\int d^5x \; \left[
- H^{\mu\nu,}{}_{\rho\sigma}\,\partial_{\lambda}D^{\rho\sigma\lambda,}{}_{\mu\nu} + {\cal{L}}(H) 
 \right ]\quad. 
\label{put3}
\end{align}
This action is invariant under the gauge transformations given by
\begin{eqnarray}
\delta H^{\mu\nu,}{}_{\rho\sigma} &=& \partial_{[\rho}\Lambda^{\mu\nu}{}_{\sigma]} + 
4\,\partial_{[\rho}\partial^{[\mu}\xi^{\nu]}{}_{\sigma]}\quad,  
\label{Hgauge} \\
\delta D^{\mu\nu\rho,}{}_{\sigma\tau} &=& -3\,\delta^{[\mu}{}_{[\sigma}\,\Lambda^{\nu\rho]}{}_{\tau]}
-12(\tfrac{3}{2}\delta^{[\mu}_{[\sigma}\partial_{\tau]}\xi^{\nu\rho]}-
\tfrac{1}{2}\delta^{[\mu}_{[\sigma}\partial^{\nu}\xi^{\rho]}{}_{\tau]}-\tfrac{3}{2}\delta^{[\mu}_{[\sigma}
\partial^{\nu}\xi_{\tau]}{}^{\rho]}+\delta^{[\mu}_{[\sigma}\delta^{\nu}_{\tau]}
\partial_\alpha\xi^{\rho]\alpha}
-\delta^{[\mu}_{[\sigma}\delta_{\tau]}^{\nu}\partial^{\rho]}
\xi_{\alpha}{}^\alpha)\,.
\nonumber \\
\label{Dgauge}
 \end{eqnarray}
It is easy to see that the gauge transformations for the $D$ field are reducible.
Indeed, $\delta_{\lambda} D^{\mu\nu\rho,}{}_{\sigma\tau} = 0$ for 
\begin{equation}
\label{reduce}
\delta \xi^{\mu\nu}=\partial^{\mu}\bar{\xi}^{\nu}\quad .
\end{equation}
It is a consequence of analogous reducibilities for $\delta_\xi H$ and 
$\delta_\xi C\,$. In addition any $\delta_\xi H$ transformation with 
antisymmetric $\xi^{\mu\nu}=-\xi^{\mu\nu}$ can be absorbed by a  
redefinition of the $\Lambda$ parameter. The same holds for $\delta_\xi D$. So the antisymmetric part of $\xi$ can be redefined away. 

Because the $D$ field enters $S[D,H]$ only through the 
quantity $\partial_{\lambda}D^{\lambda\rho\sigma,\mu\nu}$ it is obvious that 
$S[D,H]$ is invariant under the additional gauge transformation
\begin{align}
\delta_\psi D^{\rho\sigma\lambda,}{}_{\mu\nu} = 
\partial_\tau \psi^{\rho\sigma\lambda\tau,}{}_{\mu\nu} \quad ,
\end{align}
where the differential gauge parameter $\psi$ is totally antisymmetric in its two groups of indices 
separately.
\vspace*{.1cm}

Varying $S[D,H]$ with respect to $D^{\rho\sigma\lambda,}{}_{\mu\nu}$ enforces the relation 
$H_{\mu\nu,}{}^{\rho\sigma}(C)=2\,\partial^{[\rho}C_{\mu\nu,}{}^{\sigma]}\,$ that, 
when plugged back into (\ref{put3}), gives the action 
$S_{\rm Curt.}[H_{\mu\nu,}{}^{\rho\sigma}(C)]\,$. 
On the other hand, the variation of $S[D,H]$ with respect to the field
$H_{\mu\nu,}{}^{\rho\sigma}$ gives
\begin{align}
 \frac{\delta S[D,H]}{\delta H^{\mu\nu,}{}_{\rho\sigma}}\; = 
 - \partial_{\lambda}D^{\rho\sigma\lambda,}{}_{\mu\nu} + H_{\mu\nu,}{}^{\rho\sigma} 
+ 2\, H_{[\mu}{}^{[\rho,}{}_{\nu]}{}^{\sigma]} 
- 6\, \delta_{[\mu}^{[\rho}H_{\nu]\lambda,}{}^{\sigma]\lambda}
- 2\, \delta_{[\mu}^{[\rho}H^{\sigma]\lambda,}{}_{\nu]\lambda}
+ 2\, \delta_{[\mu}^{[\rho}\delta_{\nu]}^{\sigma]}H^{\alpha\beta,}{}_{\alpha\beta}\;.
\end{align}
Solving the equation $\frac{\delta S[D,H]}{\delta H^{\mu\nu,}{}_{\rho\sigma}}=0$ for 
$H^{\mu\nu,}{}_{\rho\sigma}$ gives 
\begin{eqnarray}
 H_{\mu\nu,}{}^{\rho\sigma} &=& \tfrac{1}{2}\;
( \partial_{\lambda}D^{\rho\sigma\lambda,}{}_{\mu\nu} 
    - \partial^{\lambda} D_{\mu\nu\lambda,}{}^{\rho\sigma} ) + 
\partial_{\lambda}D^{\lambda[\rho}{}_{[\mu,}{}^{\sigma]}{}_{\nu]} 
- \tfrac{3}{2}\,\partial_{\lambda}D^{\lambda\alpha[\rho,}{}_{\alpha[\mu}\,\delta_{\nu]}{}^{\sigma]}
\nonumber \\
&&+\; \tfrac{1}{2}\,
\partial^{\lambda}D_{\lambda\alpha[\mu}{}^{\alpha[\rho,}\,\delta_{\nu]}{}^{\sigma]}
+ 6\, \delta_{[\mu}^{[\rho}\delta_{\nu]}^{\sigma]}\,
\partial_{\lambda}D^{\alpha\beta\lambda,}{}_{\alpha\beta} \quad .
\label{HofD}
\end{eqnarray}
Inserting this expression back into the action $S[D,H]$ yields
\begin{equation}
S[D^{\rho\sigma\lambda,}{}_{\mu\nu}] 
= \tfrac{1}{4}\,\int d^5x \; {\cal L}(D)\quad ,
\label{ddaction5}
\end{equation}
where
\begin{eqnarray}
{\cal L}(D) &=& \left[ 
 -  \partial_{\lambda}D^{\lambda\rho\sigma,\mu\nu}\partial^{\alpha}D_{\alpha\rho\sigma,\mu\nu} 
 +  \partial_{\lambda}D^{\lambda\rho\sigma,\mu\nu}\partial^{\alpha}D_{\alpha\mu\nu,\rho\sigma} 
- 2 \,\partial_{\lambda}D^{\lambda\rho\sigma,\mu\nu}
\partial^{\alpha}D_{\alpha\mu\rho,\nu\sigma}
\right. \nonumber \\
& & \left. \; 
+\;3\,\partial_{\lambda}D^{\lambda\mu\sigma,}{}_{\nu\sigma}
     \partial^{\alpha}D_{\alpha\mu\rho,}{}^{\nu\rho}
- \partial_{\lambda}D^{\lambda\mu\sigma,}{}_{\nu\sigma}
     \partial_{\alpha} D^{\alpha\nu\rho,}{}_{\mu\rho}
-\tfrac{1}{3}\;\partial_{\lambda}D^{\lambda\mu\nu,}{}_{\mu\nu}
     \partial_{\alpha} D^{\alpha\rho\sigma,}{}_{\rho\sigma} \right ]\; . \qquad 
\label{SD}
\end{eqnarray}
We know that, by construction, the action $S[D^{\rho\sigma\lambda,}{}_{\mu\nu}]$ is equivalent
to Curtright's action $S_{\rm Curt.}[C_{\mu\nu,\rho}]$  
which in its turn is equivalent to the Fierz--Pauli action $S^{FP}[h_{\mu\nu}]\,$. 
The child action $S[D]$ inherits from its parent action $S[D,H]$ the invariance under the 
gauge transformations:
\begin{eqnarray}
\delta_{\Lambda,\xi,\psi} D^{\mu\nu\rho,}{}_{\sigma\tau} &=&  
- {3} \,\delta^{[\mu}{}_{[\sigma}\,\Lambda^{\nu\rho]}{}_{\tau]}
+  \partial_\lambda \psi^{\mu\nu\rho\lambda,}{}_{\sigma\tau}
\nonumber \\
& & -12\;(\tfrac{3}{2}\;
\delta^{[\mu}_{[\sigma}\partial_{\tau]}\xi^{\nu\rho]}-
\tfrac{1}{2}\;\delta^{[\mu}_{[\sigma}\partial^{\nu}\xi^{\rho]}{}_{\tau]}
-\tfrac{3}{2}\;\delta^{[\mu}_{[\sigma}
\partial^{\nu}\xi_{\tau]}{}^{\rho]}+\delta^{[\mu}_{[\sigma}\delta^{\nu}_{\tau]}
\partial_\alpha\xi^{\rho]\alpha}-\delta^{[\mu}_{[\sigma}\delta_{\tau]}^{\nu}\partial^{\rho]}
\xi_{\alpha}{}^\alpha)\,. 
\nonumber \\
& &
\label{transfoD}
\end{eqnarray}
%

In order to pursue the last point (iii) of the above programme, 
we decompose the $D$ field into its irreducible $\mathfrak{gl}_n$ representations
\begin{align}
D^{\mu\nu\rho},{}_{\sigma\tau} &= X^{\mu\nu\rho},{}_{\sigma\tau} + 
   \delta^{[\mu}{}_{[\sigma}\,Z^{(1)\,\nu\rho],}{}_{\tau]} + 
  \delta^{[\mu}{}_{[\sigma}\,\delta^{\nu}{}_{\tau]}\,Z^{(2)\,\rho]} 
\end{align}
and substitute this into the action $S[D]\,$. Contrary to what happens in the case of the 
Curtright action resulting from the off-shell dualisation of the Fierz--Pauli action, here 
we find that most of the components of the $Z$ fields survive in the action. 
As one can see from (\ref{transfoD}), only one component of $Z^{(1)}$ disappears from the action
(the totally antisymmetric component $Z^{(1)}{}_{[\nu\sigma,\tau]}\,$ which may be gauged away by $\Lambda_{\nu\rho\tau}$), and not all of it.
In terms of the dual $\mathfrak{gl}_5$-irreducible field $E^{(1)}{}_{\mu[3],\nu}\,$ 
corresponding to $Z^{(1)}\,$, 
(in $n$ dimensions $E^{(1)}{}_{a[n-2],b[n-4]}:=\frac{1}{(n-2)!}\; \epsilon_{c[2]a[n-2]} \;Z^{{(1)}\, c[2]},{}_{b[n-4]}\,$),  
it means that only the traceless part of $E^{(1)}$ enters the action. The field $Z^{(2)}{}_{\mu}$ also survives 
inside the action.  
The remaining fields inherit the differential gauge transformations from (\ref{transfoD}).   
We now proceed to the $n\geqslant 5$-dimensional construction. 
\vspace*{.1cm}

\paragraph*{General case with $n\geqslant 5$.}

We start from the Curtright action in dimension $n\,$, 
written in terms of 
$Y_{\mu[n-2],\nu}(C) = \partial_{\mu}C_{\mu[n-3],\nu} \,$:
\begin{equation}
\label{Curtright}
S_{\rm Curt.}[Y(C)]
=\int d^nx\; \left[Y^{\lambda[n-2],\mu}Y_{\lambda[n-2],\mu}
-\tfrac{(n-2)^2}{(n-3)}\;Y^{\lambda[n-3]\mu,}{}_\mu
Y_{\lambda[n-3]\nu,}{}^\nu+\tfrac{(n-2)}{(n-3)}\;Y^{\lambda[n-3]\rho,\mu}
Y_{\lambda[n-3]\mu,\rho}\right]
\end{equation}
and rewrite this action in terms of the following object 
\begin{equation*}
 H^{\lambda[n-3],}{}_{\mu\nu}(C) = \partial_{[\mu}C^{\lambda[n-3],}{}_{\nu]}.
\end{equation*} 
A basis $\{A_i\}_{i=1,\ldots 6 }$ of all the possible terms entering the Lagrangian is given here:
\begin{equation}
\begin{array}{rclrcl}
  A_1 & = & H^{\lambda[n-3],\mu\nu}\;H_{\lambda[n-3],\mu\nu}\quad,
\qquad 
 & 
 A_2 &=&H^{\lambda[n-4]\nu,\rho \mu}\;H_{\lambda[n-4]\rho,\nu \mu}\quad,
\\
 A_3 & =&H^{\lambda[n-5]\nu_1\nu_2,\rho_1\rho_2}\;
 H_{\lambda[n-5]\rho_1\rho_2,\nu_1\nu_2}\quad,\qquad
 &
 A_4 &=& H^{\lambda[(n-4)]\nu,}{}_{\nu}{}^\mu \;H_{\lambda[n-4]\sigma,}{}^{\sigma}{}_\mu \quad, 
\\
 A_5& =& H^{\lambda[n-5]\nu\sigma,}{}_{\sigma}{}^\rho\;  H_{\lambda[n-5]\rho\tau,}{}^{\tau}{}_\nu\quad,\qquad
 &
A_6 & =&H^{\lambda[n-5]\nu\sigma,}{}_{\nu\sigma}
\;H_{\lambda[n-5]\rho\tau,}{}^{\rho\tau}\quad .
\end{array}
\end{equation}
The resulting re-writing of Curtright's action reads
\begin{eqnarray}
S_{\rm Curt.}[H(C)] &=& \int d^n x \left( \tfrac{2}{(n-2)} A_1 +
 \tfrac{4}{(n-2)} A_2 + \bar\beta A_3 \right.
 \nonumber \\
 & & \left. \hspace*{2cm} -\; 4 A_4 -  
[\tfrac{4(n-4)}{(n-2)}\; + 4 \bar\beta] A_5
 + [ \tfrac{(n-4)(n-1)}{n-2}\; 
 + \bar\beta]A_6
\right)\quad .
\label{CurtHH}
\end{eqnarray}
Via the free parameter $\bar\beta\,$ there appears an ambiguity in the 
above form of the Lagrangian, due to the addition of total derivatives
that modify the form of the Lagrangian but do not modify the action
itself --- in the present context we discard all boundary terms. 
The linear combination $I:=A_3-4A_5+A_6$ is a total divergence and hence
does not contribute to the action. This ambiguity related to the
addition of a total derivative to the Lagrangian will be reflected in a 
one-parameter ambiguity in the resulting dual action. 

\noindent One can always rescale the action by an overall coefficient. 
After multiplying equation (\ref{CurtHH}) by $\frac{n-2}{4}$ we obtain
\begin{eqnarray}
\label{acren}
S_{\rm Curt.}[H(C)] &=&\int d^n x \Big[ 
\tfrac{1}{2}\;A_1 + A_2 + \beta A_3-(n-2)A_4
 \nonumber \\ 
& &   -\;  (n-4+4\beta) A_5 + [\tfrac{(n-1)(n-4)}4+\beta]A_6 \Big]
 ~=~\int d^nx \;{\cal L}^{\rm Curt.}(H(C)) \quad 
\end{eqnarray}
and we note that $\bar\beta = (\frac{4}{n-2})\beta$.
One recovers the action (\ref{CurtH}) by setting $n=5$ and
$\beta=0\,$. 
\vspace*{.3cm}

At this stage we view the field $H^{\mu[n-3],}{}_{\nu[2]}$ as independent and introduce a new field $D^{\mu[3],}{}_{\nu[n-3]}$ leading to the following parent action
\begin{align}
S[D^{\mu[3],}{}_{\nu[n-3]},H^{\mu[n-3],}{}_{\nu[2]}] \;=\; 
\int d^nx \; \left[
- H^{\mu[n-3],}{}_{\nu[2]}\,\partial_{\lambda}D^{\lambda\nu[2],}{}_{\mu[n-3]} + {\cal{L}}^{\rm Curt.}(H) 
 \right ]\quad. 
\label{put4}
\end{align}
The parent action $S[D^{\mu[3],}{}_{\nu[n-3]},H^{\mu[n-3],}{}_{\nu[2]}]$
is invariant under the following gauge transformations 
\begin{eqnarray}
\delta_{\Lambda,\xi} H^{\lambda[n-3],}{}_{\nu\rho} &=& 
\partial_{[\nu}\Lambda^{\lambda[n-3]}{}_{\rho]} 
+ \partial_{[\nu}\partial^{\lambda}\xi^{\lambda[n-4],}{}_{\rho]}\quad ,
\\
 \delta_{\Lambda,\xi,\psi} D_{\mu\nu\rho|}{}^{\lambda[n-3]}&=&3(-1)^{n-4}(1-2\beta)
\delta{}_{[\mu}{}^{\lambda}\Lambda{}_{\nu\rho]}{}^{\lambda[n-4]} 
 + \delta_{\xi,\psi}D_{\mu\nu\rho,}{}^{\lambda[n-3]}\quad,
 \label{deltaD}
\end{eqnarray}
where
\begin{eqnarray}
\delta_{\xi,\psi} D^{\mu[3],}{}_{\lambda[n-3]} ~=~ 3 & \!\Big( \! &  
 \gamma_1\; 
 \delta^{\mu}_{\lambda}\partial^\mu \xi_{\lambda[n-5]}{}^{\mu,}{}_{\lambda}
 +\gamma_2\;\delta^\mu_\lambda\partial^\mu\xi_{\lambda[n-4],}{}^\mu
 +
\gamma_3\;\delta_\lambda^\mu\partial_\lambda\xi_{\lambda[n-5]}{}^{\mu,\mu}+
\nonumber \\
& & \;
\gamma_4\;\delta_\lambda^\mu\partial_{\lambda}\xi_{\lambda[n-6]}{}^{\mu\mu,}{}_\lambda +\;
\gamma_5\;\delta_\lambda^\mu\delta^\mu_{\lambda}\partial^\mu\xi_{\lambda[n-5]}{}^{\nu,}{}_\nu +\;
\gamma_6\;\delta_\lambda^\mu\delta_\lambda^\mu\partial_\lambda\xi_{\lambda[n-6]}{}^\mu{}_\rho{}^{,\rho} + 
\nonumber \\
& & \;
\gamma_7\;\delta_\lambda^\mu\delta_\lambda^\mu\partial_{\rho}\xi_{\lambda[n-5]}{}^{\rho,\mu}+\;
\gamma_8\;\delta_\lambda^\mu\delta_\lambda^\mu\partial_\rho\xi_{\lambda[n-6]}{}^{\rho \mu,}{}_\lambda+\;
\gamma_9\;\delta^\mu_\lambda\delta^\mu_{\lambda}\partial_\rho\xi_{\lambda[n-5]}{}^{\mu,\rho}+\;
\nonumber \\
& & 
\gamma_{10}\;\delta_\lambda^\mu\delta_{\lambda}^\mu \delta_\lambda^\mu\partial_\rho
\xi_{\lambda[n-6]}{}^\rho{}_{\nu,}{}^{\nu}\;\Big) 
+ \partial_{\nu}\psi^{\mu[3]\nu,}{}_{\lambda[n-3]}\quad,
\label{gengautra}
\end{eqnarray}
with
\begin{align}
& \gamma_1=\tfrac{4\beta+n-4}{n-3}\;,\qquad
 \gamma_2=\tfrac{(n-2)}{(n-3)}\;,\qquad
 \gamma_5=\tfrac{2(-1)^{n-3}}{n-3}\;[\tfrac{(n-4)(n-1)}{4}\;+\beta]\;,\qquad
\nonumber \\
&
\gamma_7=-\tfrac{\gamma_3}{2}+(-1)^{n-4}\;\tfrac{(n-2)(n-4)}{2(n-3)}\;,\qquad
\gamma_8=-\gamma_4+(-1)^{n-3}(n-4+4\beta)\tfrac{(n-5)}{2(n-3)}\;,\qquad
\nonumber \\
&
\gamma_9=\tfrac{\gamma_3}{2}+(-1)^{n-4}\;\tfrac{(n-4+4\beta)}{2(n-3)}\;,\qquad
\gamma_{10}=-\tfrac{\gamma_6}{3}\; + \tfrac{2}{3}\;[\tfrac{(n-4)(n-1)}{4}\; + \beta]
\tfrac{(n-5)}{(n-3)}\;.
\end{align} 
Apart from the parameter $\beta\,$, the other free parameters are
$\{\gamma_3,\gamma_4,\gamma_6\}\,$. The freedom in the last three $\gamma$ parameters reflects
a redundancy between $\delta_{\xi}D_{\mu\nu\rho,}{}^{\lambda[n-3]}\,$ and
$\delta_{\psi}D_{\mu\nu\rho,}{}^{\lambda[n-3]}\,$.
Indeed, a $\psi$-transformation of the form
 \begin{equation}
 \psi^{\mu[4],}{}_{\lambda[n-3]}=
 \theta_1\;\delta^\mu_{\lambda}\delta^\mu_{\lambda}\;
  \xi_{\lambda[n-5]}{}^{\mu,\mu}
 +
 \theta_2\;\delta_\lambda^\mu\delta_\lambda^\mu\;\xi_{\lambda[n-6]}{}^{\mu[2],}{}_\lambda
 +
 \theta_3\;\delta^\mu_\lambda\delta^\mu_\lambda\delta^\mu_\lambda\;\xi_{\lambda[n-6]}{}^{\mu\rho,}{}_{\rho}
 \end{equation}
reproduces the 3-parameter freedom in $\{\gamma_3,\gamma_4,\gamma_6\}\,$, so that one
may keep $\psi$ arbitrary and set $\{\gamma_3,\gamma_4,\gamma_6\}\,$ to zero without 
loss of freedom in the gauge transformations. 
\vspace*{.3cm} 
 
Extremising the action (\ref{put4}) with respect to $H$ gives the 
following relation:
\begin{eqnarray}
 \partial^{\nu}D_{\mu[2]\nu,\lambda[d-3]} &=& H_1 + 2\;H_2 +
 2\beta \;H_3 - 2(n-2)(-1)^{n-4}\;H_4 
\nonumber \\
\label{var1} 
 & & - 2(n-4+4\beta)(-1)^{n-4}\;H_5+2[\tfrac{(n-1)(n-4)}{4}\;
  +\beta]\;H_6 \quad ,
\label{divDforH}
\end{eqnarray}
 where
\begin{eqnarray}
& H_1=H_{\lambda[n-3],\mu[2]}, \quad H_2=H_{\lambda[n-4]\mu,\lambda \mu}, 
 \quad H_3=H_{\lambda[n-5]\mu[2],\lambda\lambda}\quad ,&
 \nonumber \\
& H_4=\eta_{\lambda \mu}H_{\lambda[n-4]\sigma,}{}^{\sigma}{}_\mu, \quad
 H_5=\eta_{\lambda \mu}H_{\lambda[n-5]\mu\tau,}{}^{\tau}{}_\lambda, \quad
 H_6=\eta_{\lambda \mu}\eta_{\lambda \mu}
 H_{\lambda[n-5]\rho\tau,}{}^{\rho\tau}\quad .&
\label{Hi}
\end{eqnarray}
with the convention that similar indices are implicitly antisymmetrised. 
The next step amounts to inverting the equation (\ref{var1}) in 
order to express the $H$ field in terms of 
\begin{equation}
T_{\lambda[n-3],\mu[2]} := \partial^{\nu} D_{\mu[2]\nu,\lambda[n-3]}
\quad . \label{divergenceofD}
\end{equation}
Having done that, one can replace the resulting expression $H(T)$
inside the parent action in order to obtain a ``child action" 
$S[T(D)]\,$. After lengthy, but straightforward, computation 
introducing 
\begin{eqnarray}
& T_1 ~=~ T^{\lambda[n-3],\mu[2]}\;T_{\lambda[n-3],\mu[2]}\quad ,\qquad
T_2~=~T^{\lambda[n-4]\nu,\rho\mu}\;T_{\lambda[n-4]\rho,\nu \mu}\quad ,
\qquad &
\nonumber \\
&T_3 ~=~ T^{\lambda[n-5]\nu_1\nu_2,\rho_1\rho_2}\;
T_{\lambda[n-5]\rho_1\rho_2,\nu_1\nu_2}
\quad ,\qquad
T_4 ~=~ T^{\lambda[n-4]\nu,}{}_{\nu}{}^\mu \;T_{\lambda[n-4]\sigma,}{}^{\sigma}{}_\mu \quad ,\qquad &
\nonumber \\
&T_5 ~=~ T^{\lambda[n-5]\nu\sigma,}{}_{\sigma}{}^\rho \;T_{\lambda[n-3]\rho\tau,}{}^{\tau}{}_\nu \quad ,\qquad
T_6 ~=~ T^{\lambda[n-5]\nu\sigma,}{}_{\nu\sigma}\;
T_{\lambda[n-5]\rho\tau,}{}^{\rho\tau} \quad ,\qquad &
\label{notation2}
\end{eqnarray}
we find 
\begin{equation}
\label{dualaction}
S[T(D)] = -\frac{1}{2}\;\int d^n x\;
\left(a_1T_1+a_2T_2+a_3T_3+a_4T_4+a_5T_5+
a_6T_6\right)
\end{equation} 
where 
\begin{eqnarray}
 &a_1=\frac{2(n-4)(n-4-\beta (n-5))}{(1-2\beta)((n-4)(n-1)+4\beta)}
 \quad ,\qquad 
 a_2=-\frac{(n-3)((n-4)(n-7)+4\beta (n-5))}{(1-2\beta)((n-4)(n-1)+4\beta)}
 \quad , &
\nonumber \\
 & a_3 =-\frac{(n-3)(n-4)(2\beta +n-4)}{(1-2\beta)((n-4)(n-1)+4\beta)}
 \quad ,\qquad 
 a_4=-\frac{(n-3)^2(3(n-4)-4\beta (n-5))}{2(1-2\beta)
 ((n-4)(n-1)+4\beta)}\quad ,&
\nonumber \\
 & a_5=\frac{(n-3)^2(n-4)(n-4+4\beta)}{2(1-2\beta)((n-4)(n-1)+4\beta)}\quad ,\quad
 a_6=\frac{(n-3)^2(n-4)^2}{6((n-4)
 (n-1)+4\beta)}\quad . &
  \label{acoeff}
 \end{eqnarray}
Setting $n=5$ and $\beta=0$ reproduces the action 
(\ref{ddaction5})-(\ref{SD}). 
The coefficients $\{a_1\}_{i=1,\ldots 6}$ all have the same denominator, 
so one can multiply the action $S[T(D)]$ by an overall coefficient and
thereby simplify the expression for the coefficients 
$\{a_1\}_{i=1,\ldots 6}\,$.
Notice that there are singular values for $\beta\,$, \emph{i.e.}
the denominators in (\ref{acoeff}) vanish for $\beta^{(1)}=1/2$ and 
$\beta^{(2)}=\frac{-(n-4)(n-1)}{4}\;$. These values have to be rejected
since they lead to a noninvertibility in the relation between
$H$ and $T\,$.
More precisely, from (\ref{deltaD}) and $\beta=\beta^{(1)}$ 
one can see that the $D$ field loses its algebraic gauge symmetry
resulting in extra propagating degrees in freedom in $D$ compared 
to $H(C)\,$. From (\ref{divDforH})-(\ref{Hi}), 
one sees that setting $\beta=\beta^{(2)}\,$ removes the 
double-trace terms $H_6$ from the expression for the divergence of 
$D\,$ 
in terms of $H\,$. We note that the parameter $\beta$ does not exist in 
$n=4$ as there is no suitable total derivative term that may be added 
to the Fierz--Pauli Lagrangian, the latter being fixed 
unambiguously from the requirement of gauge invariance.

\vspace*{.3cm}

Before closing this section, we would like to express the dual
action $S[T(D)]$ in terms of the Hodge dual of $T\,$, 
introducing 
 \begin{eqnarray}
 \label{hdual}
 \varepsilon^{\mu[2]\nu[n-2]}U^{\lambda[n-3],}{}_{\nu[n-2]}&=& 
 T^{\lambda[n-3],\mu[2]}\quad,\quad {\rm where}
 \nonumber \\
 U^{\lambda[n-3],}{}_{\nu[n-2]}&=&
 \partial_{\nu}\widetilde{Y}_{\nu[n-3],}{}^{\lambda[n-3]}
 \quad {\rm and}
 \nonumber \\
 \widetilde{Y}_{\nu[n-3],\lambda[n-3]} &=& \tfrac{(-1)^{n-2}}{3!(n-3)!}\;
 \varepsilon_{\nu[n-3]\mu[3]}D^{\mu[3],}{}_{\lambda[n-3]}\quad .
 \end{eqnarray}
This leads to the substitution of each term in (\ref{dualaction}) 
by a corresponding group of bilinear terms in $U\,$, constructed 
analogously to the $T_i\,$, $i=1,\ldots 6\,$. 
Explicitly, 
\begin{eqnarray}
& U_1 ~=~ U^{\lambda[n-3],\mu[n-2]}\;U_{\lambda[n-3],\mu[n-2]}\quad ,\qquad
U_2~=~U^{\lambda[n-4]\nu,\rho\mu[n-3]}\;
U_{\lambda[n-4]\rho,\nu \mu[n-3]}\quad ,
\qquad &
\nonumber \\
& U_3 ~=~ U^{\lambda[n-5]\nu[2],\rho[2]\mu[n-4]}\;
U_{\lambda[n-5]\rho[2],\nu[2]\mu[n-4]}
\quad ,\qquad
U_4 ~=~ U^{\lambda[n-4]\nu,}{}_{\nu}{}^{\mu[n-3]} \;U_{\lambda[n-4]\sigma,}{}^{\sigma}{}_{\mu[n-3]} \quad ,\qquad &
\nonumber \\
&U_5 ~=~ U^{\lambda[n-5]\nu\sigma,}{}_{\sigma}{}^{\rho\mu[n-4]} \;U_{\lambda[n-3]\rho\tau,}{}^{\tau}{}_{\nu\mu[n-4]} \quad ,\qquad
U_6 ~=~ U^{\lambda[n-5]\nu\sigma,}{}_{\nu\sigma\mu[n-4]}\;
U_{\lambda[n-5]\rho\tau,}{}^{\rho\tau\mu[n-4]} \quad .\qquad &
\label{notation3}
\end{eqnarray}
The Hodge dualisation between the tensors $T$ and $U$ produces the
following transformation at the level of the bilinear terms
$T_i\,$ and $U_i\,$,  $i=1,\ldots 6\,$:
\begin{eqnarray}
&T_1 \rightarrow 2(n-2)!\,U_1\quad,\qquad
T_2 \rightarrow (n-2)!\left[U_1-(n-2)U_4\right]\quad,\qquad &
\nonumber \\
&T_3 \rightarrow (n-2)!\left[ 2U_1-4(n-2)U_4+(n-2)(n-3)U_6 \right] \quad,\qquad
T_4 \rightarrow (n-2)!\left[ U_1-(n-2)U_2 \right] \quad,\qquad &
\nonumber \\
&T_5 \rightarrow (n-2)!\left[5 U_1-(n-2)U_4-(n-2)U_2+
(d-2)(d-3)U_5 \right] \quad,\qquad &
\nonumber \\
&
T_6\rightarrow (n-2)!\,[2U_1-4(n-2)U_2+(n-2)(n-3)U_3]
\quad . &
\end{eqnarray}
Up to an overall normalisation with arbitrary parameter $\alpha\,$, 
the result is
\begin{eqnarray}
S[U(\widetilde{Y})]&=&\alpha \int d^nx \;\left\{ \Big[\tfrac{(5n^3-77n^2+387n-627)}{6(n-3)}-
\tfrac{2(n^4-20n^3+154n^2-528n+669)}{3(n-4)(n-3)}\;\beta\;\Big]U_1
\right. 
\nonumber \\
 & & \qquad\qquad +\; \Big[-\tfrac{(n-3)(n-2)(7n-37)}{6}\; 
 + 
 \tfrac{2(n-3)(n-2)(2n^2-22n+59)}{3(n-4)}\;\beta\;\Big]U_2
\nonumber \\
 & & \qquad\qquad+\; 
\Big[\tfrac{(n-4)(n-3)^2(n-2)}{6}\;-
\tfrac{(n-4)(n-3)^2(n-2)}{3}\; {\beta}\; \Big]U_3
\nonumber \\
 & & \qquad\qquad+\; 
 \Big[-\tfrac{(n-2)(n^2-17n+58)}{2} \;- \tfrac{2(n-2)(n^2-13n+38)}{n-4}\;\beta\;\Big]U_4
\nonumber \\
 & & \qquad\qquad+\; 
 \Big[\tfrac{(n-4)(n-3)^2(n-2)}{2}\;+2 (n-3)^2(n-2)\beta\;\Big]U_5
\nonumber \\
 & & \qquad\qquad +\; \left. 
 \Big[-\;(n-4)(n-3)(n-2)-2 (n-3)(n-2)\beta\;\Big]U_6 \right\} 
 \quad ,
 \label{fin1}
 \end{eqnarray}
where we have made explicit the freedom in overall normalisation 
with coefficient $\alpha\,$. The action is valid for $n>4\,$. 
For $n=4$ all the procedure can be reproduced and as 
expected we find, up to a rescaling, the Fierz--Pauli action 
(without a $\beta$ parameter) where only the terms in $U_1$, $U_2$ and 
$U_4$ remain. 

The following change of variables slightly simplifies the result: 
 \begin{eqnarray}
 &\alpha = \frac{B+2\frac{A}{n-4}}{(n-3)(n-2)}, \quad 
 \alpha\beta =\frac{B-A(n-1)}{2(n-3)(n-2)}\quad ,\quad 
 i.e. &
\nonumber \\ 
 & A=\alpha(1-2\,\beta)(n-4), \quad B = 4\,\alpha\left[ \tfrac{(n-4)(n-1)}{4}\; + \beta\right] \quad .&
\end{eqnarray}
The variables $A$ and $B$ are chosen because they appear in the
denominators of (\ref{acoeff}).
In terms of these variables the action (\ref{fin1}) acquires the 
following form
\begin{eqnarray}
S[U(\widetilde{Y})]&=&\int d^n x\; \Big[ 
(\tfrac{n^2+11n-36}{3(n-2)}\;A+\tfrac{(n-5)(n^2-11n+26)}{2(n-4)(n-3)
(n-2)}\; B)U_1 
\nonumber \\
& & +\; 
(-\tfrac{2}{3}\;(n-6)(n-2)A-\tfrac{(n-5)(n-2)}{2(n-4)}\;B)U_2
\nonumber \\
& & +\; 
(\tfrac{1}{6}\;(n-3)^2(n-2)A)U_3
\nonumber \\
& & +\; 
((n-8)A-\tfrac{n^2-16n+52}{2(n-4)}\;B) U_4+
\nonumber \\
& & +\; 
(-(n-3)(n-2)A+\frac{1}{2}(n-3)(n-2)B)U_5+
\nonumber \\ 
& &  +\; ((n-3)A-(n-3)B)U_6 \Big] \quad .
\label{fin2}
\end{eqnarray}

\section{Infinitely many off-shell dualisations}
 \label{sec:Mtimes}

We have seen that the double-dual formulation of Fierz--Pauli 
gravity is less economical than the original formulation or
than Curtright's formulation in the sense that a larger spectrum
of fields is needed for the manifestly covariant and local action. 

In this section we show that one can actually describe linearised
gravity around a flat background in infinitely many dual ways, 
each being manifestly Poincar\'e covariant and local, but featuring
more and more fields. 
 
\paragraph{Tower based on the Fierz--Pauli field.} 

Starting from the Fierz--Pauli action 
\begin{eqnarray}
S[h_{[1,1]}] = \int d^ nx \;L^{\rm FP}(\partial_{\alpha}h_{\mu,\nu})=\int d^ nx\;  \partial^\alpha 
h^{\mu,\nu}\,\partial_\alpha h_{\mu,\nu}+\ldots\,, 
\end{eqnarray}
where 
$h_{[1,1]}\sim \tiny\yng(1)\otimes \tiny\yng(1)\,$, one introduces 
the independent field $G_1^{\alpha,\mu,\nu}$ which transforms in 
the representation 
$\tiny\yng(1)\otimes \tiny\yng(1)\otimes \tiny\yng(1)\,$
of $\mathfrak{gl}_n$ contrary to the curl 
$\Omega \sim \partial_{[\alpha}h_{\mu],\nu}\sim \tiny\yng(1,1)\otimes \tiny\yng(1)$ that enters the linearisation of the action (\ref{EH})
and from which one arrives at the Curtright action via off-shell 
Hodge duality. 
One then writes the parent action 
\begin{eqnarray}
S_{\rm FP}^{(P1)}[G_1,F_1] = 
\int d^ nx \; \left( G_{\alpha,\mu,\nu} \partial_{\beta}F^{\beta\alpha,\mu,\nu}
- \tfrac{1}{2}\; L^{\rm FP}(G_1)\right)\quad,
\end{eqnarray}
where $F_1\sim\tiny\yng(1,1)\otimes \tiny\yng(1)\otimes \tiny\yng(1)\,$.

Repeating the procedure used in the previous sections, from that parent
action one either reproduces the Fierz--Pauli action 
$S_{\rm FP}[h_{[1,1]}]$
upon extremising with respect to $F_1$ or another equivalent action 
\begin{eqnarray}
S_{\rm FP}^{(1)}[{h}^{(1)}_{[n-2,1,1]}] = \int d^nx\; 
\left[\partial_{[\mu}{h}^{(1)}{}_{\mu[n-2]],\nu,\rho}\;
\partial^{[\mu}{h}^{(1)}{}^{\mu[n-2]],\nu,\rho} + \ldots\right]\quad,
\end{eqnarray}
expressed in terms of the field
$h^{(1)}_{[n-2,1,1]}$ obtained by Hodge dualising 
$F_1$ on the first column.
For example, in dimension $n=5$ the action $S_{\rm FP}^{(1)}$ will feature the reducible field ${h}^{(1)}_{[3,1,1]}$ that decomposes 
under $\mathfrak{gl}_5$ into the following fields 
\begin{eqnarray}
\Yvcentermath1\tiny\yng(1,1,1)\quad\otimes\quad \Yvcentermath1 \tiny\yng(1)\quad \otimes \quad \Yvcentermath1\tiny\yng(1) 
\quad\quad~\sim~\quad\quad
 \underbrace{\Yvcentermath1 \tiny\yng(3,1,1)}_{\tilde{h}^{(1)}}\quad\oplus \quad \tiny\yng(2,2,1)\quad\oplus\quad2\times \tiny\yng(2,1,1,1)\oplus \quad \tiny\yng(1,1,1,1,1)\qquad .
 \label{h1}
\end{eqnarray}
Several comments are in order. 

Firstly, the child action $S_{\rm FP}^{(1)}$ inherits gauge invariances 
from its parent. 
The latter possesses an extension of the gauge invariances of the
original Fierz--Pauli action. In particular, the field 
$h^{(1)}_{[n-2,1,1]}$ will be invariant under an algebraic gauge
transformation containing an antisymmetric rank-2 tensor. In other
words, a small set of fields in (\ref{h1}) will be gauged away. 

Secondly, by construction we know that the child action $S_{\rm FP}^{(1)}$ 
propagates the same physical on-shell degrees of freedom
as the original Fierz--Pauli action, and we anticipate, drawing
from our experience with Curtright's action and with the double-dual
action, that the on-shell field, in the light-cone gauge, will be
given by the first field entering the decomposition of (\ref{h1}) as
\begin{eqnarray}
\tilde{h}^{(1)}_{i[n-2],k,l} &\approx & \epsilon_{i[n-2]} h_{kl}
\quad, 
\end{eqnarray}
where $h_{kl}$ is the $\mathfrak{so}_{n-2}$ on-shell physical graviton. 
Clearly, $\tilde{h}^{(1)}_{i[n-2],k,l}$ is not traceless, and in this
sense is not similar to $h_{kl}\,$. 
But although the field $\tilde{h}^{(1)}_{i[n-2],k,l}$ is not traceless, 
it is nevertheless non-identically vanishing and is propagating. 
The field $\tilde{h}^{(1)}_{i[n-2],k,l}$ transforms in exactly the same 
irreducible $\mathfrak{so}_{n-2}$ representation as does $h_{kl}\,$, 
namely the spin-2 representation, and therefore gives yet another 
dual formulation of the graviton, like Curtright's and the double-dual
formulations in $n$ dimensions that we have examined previously.  

Analogously, we anticipate that further dual off-shell formulations
of Fierz--Pauli theory will be given by an infinite number of actions
$S^{(m)}[h^{(m)}_{[n-2,n-2,\ldots,n-2,1,1]}]$ for $m=2,3,\ldots$, 
where the gauge field  $h^{(m)}$ possesses $m$ sets of $n-2$ 
antisymmetric indices on top of the two indices $\mu,\nu$ carried
by the original Fierz--Pauli field $h_{\mu,\nu}\,$. In particular, 
the field will contain the $\mathfrak{gl}_n$-irreducible component with the
following symmetry type
\begin{equation}
\label{FPtower}
\tilde{h}^{(m)}\qquad \sim \qquad 
\ytableausetup
{mathmode, boxsize=1.4em}
\begin{ytableau}
\mbox{{\scriptsize n}} & \mbox{{\scriptsize n}} & \none[\dots] & \mbox{{\scriptsize n}}
& \mbox{{\scriptsize n}} &\mbox{{\scriptsize n}} \\
\mbox{{\scriptsize n-1}} & \mbox{{\scriptsize n-1}}  & \none[\dots]& \mbox{{\scriptsize n-1}}  \\
\none[\vdots] & \none[\vdots]
& \none[\ldots] & \none[\vdots]  \\
\scriptstyle 4 & \scriptstyle 4 & \none[\dots]& \scriptstyle 4 
 \\
\scriptstyle 3 & \scriptstyle 3 & \none[\dots]
& \scriptstyle 3 \\
\end{ytableau}\quad. 
\end{equation}

In order to see this, one starts from the resulting child action 
$S_{\rm FP}^{(1)}[{h}^{(1)}_{[n-2,1,1]}]\,$, and notes that the
basic object entering the Lagrangian is the gradient of ${h}^{(1)}$ 
and not its curl on the first column (integrating by parts can ``undo''
the anti-symmetrisations appearing in the curl). Denote the resulting
gradient by the symbol $G_2$ with the symmetry type 
$[n-2]\otimes [1]\otimes [1]\otimes [1]\,$. 
A parent action is then obtained which features $G_2$ viewed
as an independent field together with a new field $F_2$ with the
symmetry type 
$[n-2]\otimes[2]\otimes [1]\otimes [1]\,$.
extremising the parent action with respect to $G_2$ and substituting the solution of the resulting algebraic equation inside the parent 
action will produce the child action
$S_{\rm FP}^{(2)}[{h}^{(2)}_{[n-2,n-2,1,1]}]$ in terms of the 
gauge field ${h}^{(2)}_{[n-2,n-2,1,1]}$ obtained from $F_2$ by 
Hodge dualising the second column.
Again, on-shell, the physical degrees of freedom will be carried by the
component 
$\tilde{h}^{(2)}_{i[n-2],j[n-2],k,l}$ equivalent to $h_{kl}$
via the relation 
\begin{eqnarray}
\tilde{h}^{(2)}_{i[n-2],j[n-2],k,l}\propto 
\epsilon_{i[n-2]}\,\epsilon_{j[n-2]}h_{kl}\quad.
\end{eqnarray}
Again, the resulting action 
$S_{\rm FP}^{(2)}[{h}^{(2)}_{[n-2,n-2,1,1]}]$ can be dualised
to give $S_{\rm FP}^{(3)}[{h}^{(3)}_{[n-2,n-2,n-2,1,1]}]\,$ and 
so on and so forth, each one containing the $\mathfrak{gl}_n$-irreducible 
field $\tilde{h}^{(m)}$ depicted in (\ref{FPtower}),
for $m=1,2,3,\ldots$

\paragraph{Dual graviton tower.}

In exactly the same way as we did starting from Fierz--Pauli's action, 
one can now start from Curtright's action and produce the tower of
Hodge-dual actions 
$S_{\rm Curt.}^{(m)}[{C}^{(m)}_{[n-2,\ldots,n-2,n-3,1]}]$ that
will each propagate the gauge field 
$\tilde{C}^{(m)}_{[n-2,\ldots,n-2,n-3,1]}]$ with 
$\mathfrak{gl}_n$-irreducible symmetry depicted as follows:
\begin{equation}
\label{Curttower}
\tilde{C}^{(m)}\qquad \sim \qquad 
\ytableausetup
{mathmode, boxsize=1.4em}
\begin{ytableau}
\mbox{{\scriptsize n}} & \mbox{{\scriptsize n}} & \none[\dots] & \mbox{{\scriptsize n}}
& \mbox{{\scriptsize n}} & \mbox{{\scriptsize n}}\\
\mbox{{\scriptsize n-1}} & \mbox{{\scriptsize n-1}}  & \none[\dots]& \mbox{{\scriptsize n-1}} & \mbox{{\scriptsize n-1}}  \\
\none[\vdots] & \none[\vdots]
& \none[\ldots] & \none[\vdots]  & \none[\vdots]  \\
\scriptstyle 4 & \scriptstyle 4 & \none[\dots]& \scriptstyle 4& \scriptstyle 4 
 \\
\scriptstyle 3 & \scriptstyle 3 & \none[\dots]
& \scriptstyle 3 \\
\end{ytableau}\quad. 
\end{equation}
where the number of columns with length $(n-2)$ is $m\,$. 

On-shell, all these fields will be equivalent to the Curtright field, 
which is itself equivalent to the Fierz--Pauli field in the appropriate
spacetime dimension $n\,$. In other words, all these fields, on-shell, 
transform in the spin-2 representation $\tiny\yng(2)$ of 
$\mathfrak{so}_{n-2}\,$. The corresponding actions 
$S^{(m)}_{\rm Curt.}$ with $m=1,2,\ldots $ 
give all different dual formulations of the same Fierz--Pauli action.  

In the case $m=1\,$ in five dimensions, the off-shell field is
${C}^{(1)}_{[3,2,1]}$ and decomposes under $\mathfrak{gl}_5$ into the following fields 
\begin{eqnarray}
\Yvcentermath1\tiny\yng(1,1,1)\quad\otimes\quad \tiny\yng(1,1)\quad\otimes\quad \tiny\yng(1) 
\quad\quad~\sim~\quad\quad
 \underbrace{\tiny\yng(3,2,1)}_{\tilde{C}^{(1)}}\quad\oplus \quad \tiny\yng(2,2,2) \quad\oplus\quad \tiny\yng(3,1,1,1)\quad\oplus\quad 2\times\tiny\yng(2,2,1,1)\quad \oplus\quad  2\times \tiny\yng(1)\qquad .
 \label{C1}
\end{eqnarray}

\paragraph{The double-dual's tower.} 

Finally, the same analysis can be done based on the double-dual action
given in Section \ref{sec:Lagrangian} to produce the tower of dual 
actions $S^{(m)}_{\rm dd}[D^{(m)}]$ with $m=1,2,\ldots $ that 
will each propagate the gauge field 
$\tilde{D}^{(m)}_{[n-2,\ldots,n-2,n-3,n-3]}$ with 
$\mathfrak{gl}_n$-irreducible symmetry depicted as follows:
\begin{equation}
\label{Doubledualtower}
\tilde{D}^{(m)}\qquad \sim \qquad 
\ytableausetup
{mathmode, boxsize=1.4em}
\begin{ytableau}
\mbox{{\scriptsize n}} & \mbox{{\scriptsize n}} & \none[\dots] & \mbox{{\scriptsize n}}
& \mbox{{\scriptsize n}} & \mbox{{\scriptsize n}} \\
\mbox{{\scriptsize n-1}} & \mbox{{\scriptsize n-1}}  & \none[\dots]& \mbox{{\scriptsize n-1}} & \mbox{{\scriptsize n-1}}& \mbox{{\scriptsize n-1}}  \\
\none[\vdots] & \none[\vdots]
& \none[\ldots] & \none[\vdots]  & \none[\vdots] & \none[\vdots]  \\
\scriptstyle 4 & \scriptstyle 4 & \none[\dots]& \scriptstyle 4
& \scriptstyle 4 & \scriptstyle 4 
 \\
\scriptstyle 3 & \scriptstyle 3 & \none[\dots]
& \scriptstyle 3 \\
\end{ytableau}\quad. 
\end{equation}
where the number of columns with length $(n-2)$ is $m\,$.

\section{Maximal Supergravity, Dual Gravity and $E_{11}$}
\label{sec:E11 notes}
In the previous sections we have constructed manifestly, and in outline, the actions for an infinite set of dual formulations of linearised gravity. These dual formulations and indeed the considerations that underly their construction form part of the striking story of $E_{11}\,$, the conjectured symmetry algebra of M-theory \cite{West:2001as}. 
The dual graviton tower of fields contained within $E_{11}$ and argued to be dual descriptions of gravity in \cite{Riccioni:2006az} have been shown to be equivalent to linearised gravity at the level of the action. In addition to the fields in the dual graviton tower a set of supplementary mixed-symmetry fields will appear in the action, see equation (\ref{Curttower}) where the five-dimensional supplementary fields are shown for the first field in the dual graviton tower. We will show that all the fields required to construct the actions for each
of the individual fields entering the dual graviton tower are all contained within $E_{11}$ and are associated with null and imaginary roots.

The work in this paper was inspired, in part, by the work of Hull on the double-dual graviton \cite{Hull:2000zn,Hull:2001iu,Hull:2000rr} and we commence this section with a search of the fields of $E_{11}$ seeking the double-dual graviton. The double-dual graviton was identified by Hull \cite{Hull:2000zn} within the strongly 
coupled sector of five-dimensional maximally supersymmetric supergravity. 
In this section we will identify within the low levels of $E_{11}$ the 
bosonic multiplets of maximal supergravity in five dimensions and the lift 
of these degrees of freedom into the $n=6$ multiplets again contained within 
$E_{11}\,$. The multiplets of maximal supergravity in five and six dimensions derived from $n$ and $n-1$ forms have been found using $E_{11}$ in \cite{Riccioni:2007au,Riccioni:2009xr} and using the very extension of real forms of $E_8$ in \cite{Riccioni:2008jz}. We will see that it is not possible to identify in a straightforward 
way the six dimensional $(4,0)$ multiplets within $E_{11}$ that were originally found in \cite{Strathdee:1987p5523} and which include the double-dual graviton. 
However 
each dual graviton in the dual gravity tower of fields carries the same number of degrees of freedom as the double-dual graviton and these fields do appear naturally within the decomposition of $E_{11}$ together with the supplementary fields required to construct the actions described in this paper.   

\subsection{${\cal N}_5=8$ maximal supergravity}

The maximal supergravity in 5D, having ${\cal N}=8\,$, 
may be decomposed into representations of the little group in 5D 
$Spin(3)$ and representations of $Sp(4)\,$, which is the local symmetry 
of the discrete U-duality group $\frac{E_6}{Sp(4)}\;$. 
The on-shell multiplet splits into
\begin{equation}
({\bf 1},{\bf{42}})\oplus ({\bf 2},{\bf 48}) \oplus ({\bf 3},{\bf 27}) \oplus ({\bf 4},{\bf 8})\oplus ({\bf 5}, {\bf 1})
\end{equation}
giving $2^8$ degrees of freedom. 
The bosonic degrees of freedom are given by {\bf 42} scalars, {\bf 27} vectors and 
{\bf 1} graviton (a symmetric 2-tensor, or bivector, field) in five dimensions. 
The decomposition of $E_{11}$ gauge fields at low levels 
quickly identifies these U-duality multiplets for the bosonic fields. 
Consider the Dynkin diagram for $E_{11}$
\begin{equation}
\nonumber \includegraphics[scale=0.6,angle=0]{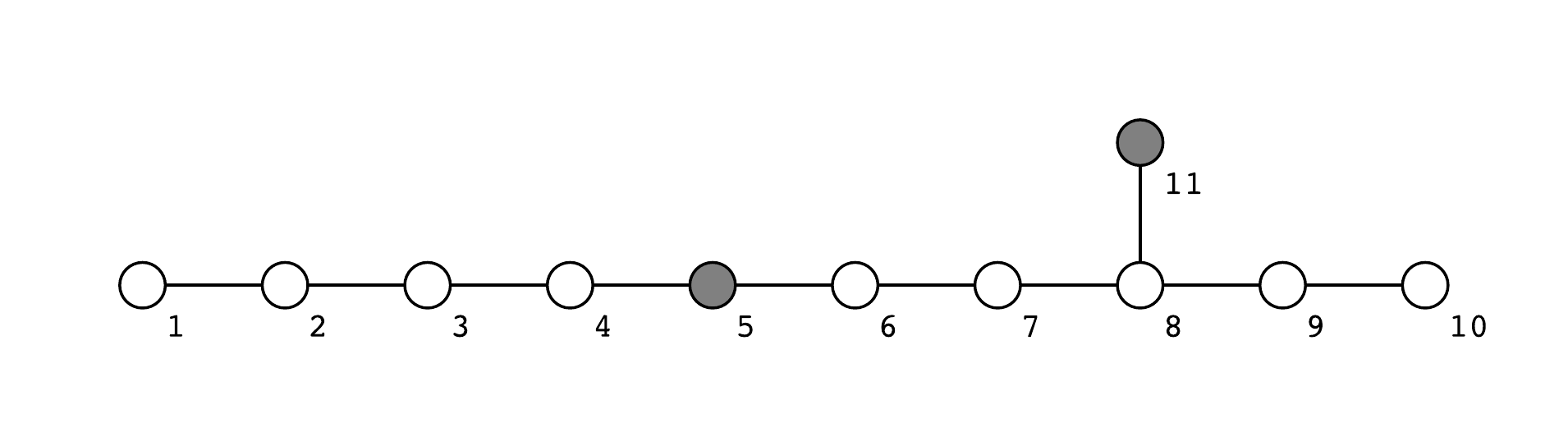}
\end{equation}
where the shaded nodes indicate the decomposition relevant to the five-dimensional theory. 
The nodes $1,\ldots 4$ make up the Dynkin diagram of $A_4$, or $\mathfrak{sl}_{5}$, 
whose non-compact sub-group $SO(1,4)$ will be the local Lorentz group for the five-dimensional spacetime. 
The remaining nodes $6,\ldots 11$ give the Dynkin diagram of $E_6\,$, 
which contains the U-duality group in five dimensions 
and we will refer to this as the internal symmetry in five dimensions.

The positive roots of $E_{11}$ may be written as a sum of the simple positive roots 
$\vec \alpha_i$ for $i=1,2,\ldots 11$ and will have the generic form:
\begin{equation}
\vec\beta= \sum_{i=1}^{11} m_i \vec \alpha_i \;.
\end{equation}
The simple root associated with node 5 may be split into a vector 
in the $A_4$ weight lattice, a vector in the $E_6$ weight lattice 
and a part which is orthogonal to the fundamental weights of both $A_4$ and $E_6\,$. 
We have 
\begin{equation}
\alpha_5=-\lambda_4+x-\nu_6 \label{alpha5}
\end{equation}
where $\lambda_i$ for $i\in\{1,2,3,4\}$ are the four fundamental weights of $A_4$ (indicated by nodes 1 to 4), 
$\nu_I$ for $I\in\{6,7,\ldots,11\}$ is a fundamental weight of $E_6$ (nodes 6 to 11) and $x$ is a vector in the weight 
lattice of $E_{11}$ but orthogonal to the weight lattices of $A_4$ and $E_6\,$. 
The decomposition of $\alpha_5$ in equation (\ref{alpha5}) guarantees that the inner products of the simple roots of 
$E_{11}$ are preserved, \emph{i.e.} $\langle\alpha_5,\alpha_i\rangle = -\delta_{i4}\,$, 
$\langle\alpha_5,\alpha_I\rangle=-\delta_{I6}\,$ and 
$\langle\alpha_5,\alpha_5\rangle=2\,$. 
The last condition is used to normalise $x\,$. 

Deletion of node 5 of the $E_{11}$ Dynkin diagram splits the roots of $E_{11}$ 
into a lowest weight representation of $A_4$ with Dynkin labels $p_i\,$, 
\begin{equation}
-\sum_{i=1}^4p_i\lambda_i=-m_5\lambda_4+\sum_{i=1}^4 m_i\alpha_i \label{A4weights}
\end{equation}
and a lowest weight representation of $E_6$ with Dynkin labels $q_I$
\begin{equation}
-\sum_{I=6}^{11}q_I\nu_I=-m_5\nu_6+\sum_{I=6}^{11} m_I\alpha_I.
\end{equation}
The coefficients $p_i$ and $q_I$ label lowest weight representations of $A_4$ and $E_6$ respectively. 
It is useful to further decompose the $E_6$ highest weight 
representation into representations of its $A_5$ sub-algebra, indicated by nodes 6 to 10, 
so that the generators of the $E_6$ algebra may be written as $A_5$ tensors. 
This is achieved by deleting the simple root associated with node 11, 
which itself decomposes into a vector in the $A_5$ weight lattice ($-\mu_8$) 
and a vector orthogonal ($y$) as $\alpha_{11}=-\mu_8+y\,$, 
where $\mu_{\hat{J}}$ for $\hat{J}\in\{6,7,\ldots 10\}$ are the fundamental weights of $A_5\,$. 
Consequently we have weights of $A_5\,$, corresponding to the decomposed (internal) $E_6\,$, 
such that 
\begin{equation}
-\sum_{\hat{J}=6}^{10}r_{\hat{J}}\mu_{\hat{J}}=-m_5\mu_6-m_{11}\mu_8+\sum_{\hat{J}=6}^{10} m_{\hat{J}}\alpha_{\hat{J}}\;.
\label{A5weights}
\end{equation}
By taking inner products with $\alpha_j$ ($j\in\{1,2,3,4\}$) in (\ref{A4weights}) we find formulae for the 
coefficients $p_j$ that label the lowest weight representations of $A_4$: 
\begin{align}
p_1&=-2m_1+m_2,\\
p_2&=m_1-2m_2+m_3,\\
p_3&=+m_2-2m_3+m_4 \quad \mbox{and}\\
p_4&=+m_3-2m_4+m_5.
\end{align}
While by taking inner products with $\mu_{k+5}$ where $k\in\{1,2,3,4,5\}$ in equation (\ref{A5weights}) 
we find formulae for the coefficients $m_{k+5}$ that label the root string 
subtracted from the highest weight representation of $A_5$:
\begin{align}
m_{k+5}&=-\sum_{j=1}^{5}r_{(j+5)}\frac{j(6-k)}{6}+m_5\frac{(6-k)}{6}+m_{11}\frac{3(6-k)}{6}\\
&=N+m_5+3m_{11}-(6-k)\sum_{j=1}^5 r_{(j+5)}-\frac{k}{6}(N+m_5+3m_{11})\in\mathbb{Z}^+ \label{mk}
\end{align}
where $N\equiv \sum (6-j)r_{(j+5)}$ is the number of indices on the tensor representation of $A_5\,$. 

The bosonic content of the supermultiplet in five dimensions can be quickly reconstructed from 
$E_{11}$ using the formulae for $p_i$ and $r_i$ above. We are interested in the scalar, vector and 
symmetric two-tensor representations of $A_4$ which correspond to $p_4=0,1,2$ respectively and 
$p_1=p_2=p_3=0\,$. 
{}From the formulae above we see that this implies $m_2=2m_1\,$, $m_3=3m_1\,$, $m_4=4m_1$ 
and hence $p_4=-5m_1+m_5\,$. 
So for the scalar multiplet we have $p_4=0$ which is satisfied by the pairs 
$(m_1,m_5)=\{(0,0),(1,5),(2,10),\ldots \}\,$. 
We will find that the scalar multiplet of five-dimensional maximal supergravity 
is found within the first pair $m_1=m_2=m_3=m_4=m_5=0\,$. 
Without any loss of generality we find from (\ref{mk}) that $N=3m_{11}-m_5\geqslant0$ 
guarantees that $m_{k+5}\in \mathbb{Z}^+\,$. 
When $m_5=0$ we have $N=3m_{11}\,$ and 
we can then identify generators in $A_5$ associated with the scalar multiplet in $A_4$ for $m_{11}=1,2,3\,$. Using
\begin{align}
m_{k+5}=(6-k)(m_{11}-\sum r_{(j+5)})\geqslant 0
\end{align}
we observe that $\sum r_{(j+5)} \geqslant  m_{11}\,$. 
When $m_{11}=0$ then $N=0$ and $\sum r_{(j+5)} = 0$   
so we find the Cartan sub-algebra ${K^M}_M$ ({\bf 6}) 
and the positive generators of $A_5$ (${K^M}_N$) for $N>M$ ({\bf 15}). 
When $m_{11}=1$ then $N=3=\sum(6-j)r_{j+5}$ and $\sum r_{(j+5)} \leqslant 1$ is satisfied by $r_8=1$ and all other $r_j=0$ 
which corresponds to a three-form $R^{M[3]}$ ({\bf 20}). 
Finally when $m_{11}=2$ we have $N=6$ which is satisfied by $r_6=1$ 
while the remaining $r_j=0\,$. 
This gives a six-form generator $R^{M[6]}$ ({\bf 1}). 
All other possibilities for sets of $r_j$ are ruled out as the associated 
root string in $A_5$ has length squared greater than two. 
This completes the scalar multiplet having dimension {\bf 42}
\begin{equation}
\phi\equiv \{ {K^M}_M({\bf 6}), {K^M}_N({\bf 15}), R^{M[3]}({\bf 20}), R^{M[6]}({\bf 1})\}
\end{equation}
where $M_i \in\{1,2,\ldots 6\}$ are internal $A_5$ tensor indices. 

The origin of the five-dimensional field content is obvious from the low level generators of $E_{11}\,$. 
The low level $E_{11}$ generators are
\begin{equation}
{K^A}_A,\, {K^A}_B ,\, R^{A[3]},\, R^{A[6]},\, R^{A[8],B}, \ldots
\end{equation}
where the $A$ and $B$ indices are eleven-dimensional. One may quickly find the scalar multiplet of five dimensional 
supergravity when the dimensionally reduced generators have only internal $E_6$ indices. 
Since these internal indices are six-dimensional the $R^{A[8],B}$ generator, 
as well as other higher level generators, do not contribute the scalar multiplet in five dimensions. 

While one could repeat the derivation of the five-dimensional scalar multiplet as initially outlined above to find the 
vector multiplet it is much simpler to achieve the same end by the method of partitioning the indices of the low level 
generators of $E_{11}$ into internal and worldvolume indices. For the vector multiplet it will suffice to dimensionally 
reduce these fields so that only one of the reduced indices is a five-dimensional world-volume index. This corresponds to $p_4=1$ and hence $m_1=0,m_5=1\,$. 
The only possibilities amongst the low level generators are
\begin{equation}
\phi^\mu\equiv \{ {K^\mu}_N ({\bf 6}), R^{\mu M[2]} ({\bf 15}), R^{\mu M[5]}({\bf 6})\}
\end{equation}
where we indicate in brackets the dimension of the internal $A_5$ tensor representation. 
In total we find the ${\bf 27}$ of the vector multiplet. 

The gravitational degrees of freedom are contained in the $(\bf 5,\bf 1)$ of $Spin(3)\otimes E_6$ which corresponds to a symmetric rank two space-time tensor carrying a trivial representation of $E_6\,$. 
This representation corresponds to $p_4=2$ and hence $m_1=0$ and $m_5=2\,$. 
However there is no such representation appearing directly in the decomposition of $E_{11}\,$. 
Instead we may identify the vielbein field from which we can construct the graviton, 
or we may pursue an alternative path to understand the origin of the gravitational degrees of freedom 
within higher level $E_{11}$ generators. 
In the first instance we may identify the traceless part of the Borel sub-algebra of $A_4$ as the $\bf 5$ 
of the little group in five-dimensions
\begin{equation}
\phi^{(\mu\nu)}_{\mbox{Gravity}}=\{{K^\mu}_\nu ({\bf 5},{\bf 1}) \} \qquad \mbox{where } 
\nu\geqslant\mu
\end{equation}
where we have indicated in brackets the $A_4\otimes E_6$ multiplet. 
The fields associated with these generators, ${h_\nu}^\mu$ give the vielbein 
${e_\nu}^\mu={(e^{-h})_\nu}^\mu$ from which the graviton may be reconstructed, 
see section two of \cite{Englert:2003p595} for a detailed discussion of the vielbein within $E_{11}$. 

Amongst the low level generators of $E_{11}$ there is a second way the gravitational degrees of freedom 
may be identified using the dual graviton. Upon dimensional reduction we find a singlet of $A_5$ containing 
the five-dimensional dual graviton $R^{\mu[2]M[6],\nu}\,$. 
We note that the traceless part of $\tiny\yng(2,1)$ as a representation of the little group in five dimensions 
also gives the ${\bf 5}\,$. 
We may dualise the dual graviton at the linear level in its two antisymmetrised indices to a symmetric 
two tensor and a singlet of the internal $A_5$:
\begin{equation}
\phi^{(\mu\nu)}_{\mbox{Dual gravity}}=\{\star_1{R}^{\mu[2] M[6],\nu}\}
\end{equation}
where $\star_1 = \iota_1 \star_H d_1$ is the action on the first set of two antisymmetric indices associated with the Hodge dual $\star_H$ in five dimensions, $d$ is the exterior derivative, $\iota$ is the interior product; 
the  index on $\star$, $\iota$ and $d$ indicates the set of antisymmetric indices of the mixed-symmetry tensor that the operations acts on - it indicates the column as numbered from left to right on the associated mixed-symmetry Young 
tableau. 
This is not the end of the story as amongst all the generators of $E_{11}$ there is an infinite tower 
of generators each of which alone may encode the ${\bf 5}$ gravitational sector in a similar fashion to the 
dual graviton. 
These are fields which we would naturally associate with duals of the graviton, their presence was highlighted in \cite{Damour:2002cu,Riccioni:2006az}  
and the first of these we would suspect to be linked to Hull's double dual graviton. In five dimensions, the dual graviton $c_{\mu[2],\nu}$ and the double dual graviton 
$d_{\mu[2],\nu[2]}$ are related to the graviton $h_{\mu\nu}$ by 
$c_{\mu[2],\nu}=\star_1 h_{\mu\nu}$ and $d_{\mu[2],\nu[2]}=\star_2 c_{\mu[2],\nu}=\star_1\star_2 h_{\mu\nu}\,$, 
the set of fields that occur naturally in $E_{11}$ is associated with 
$\star_3 \star_1 h_{\mu\nu}=\bar{c}_{\lambda[3],\mu[2],\nu}$ that is the dual of a trivial ``scalar''. In fact we recognise these fields as the 
dual tower whose Young tableau are shown in (\ref{Curttower}), and we understand from section 2 the Curtright action may be reconciled with the double-dual action via a parent action.
The relevant higher level generators of $E_{11}$ written as eleven-dimensional tensors have the form 
\begin{equation}
R^{A^{(1)}[9]A^{(2)}[9] \ldots A^{(m)}[9]B[8]C} \label{11Dmultidualgraviton}
\end{equation}
for all $m\geqslant 0$\footnote{when $m=0$ we find the dual graviton.}.  
Upon reduction to five dimensions these generators are parameterised by fields 
which might also be interpreted as dual gravitational fields having the form
\begin{equation}
\phi^{(\mu\nu)}_{\mbox{(m+1)-dual gravity}}=\{{R}^{\mu^{(1)}[3] \mu^{(2)}[3] \ldots \mu^{(m)}[3]\nu[2]\rho}\} 
\label{5Dmultidualgraviton}
\end{equation}
where the internal indices, which have been suppressed, 
carry the singlet representation of the internal $E_6\,$. 
The $m$ sets of antisymmetric $\{\mu[3]\}$ indices transform 
trivially as a representation of the little group in five dimensions. 
Hence each of these generators is associated with fields carrying 
the same degrees of freedom as the traceless part of $\tiny\yng(2,1)$, 
the dual graviton in five dimensions. 
We will return to discuss this set of fields in the sequel.
%
\subsection{Six-dimensional theories from $E_{11}$.}
%
The usual lift of the five-dimensional multiplet of scalars, vectors and bivectors 
to six dimensions traces their origin to six-dimensional tensor objects of the same and higher rank. 
The setting of $E_{11}$ does nothing to change this, but it will be useful to explicitly reproduce 
the lifting of the five-dimensional multiplet. 
The six-dimensional field content is reproduced from $E_{11}$ by deleting nodes 6 and 11 
on the $E_{11}$ diagram as indicated below.
\begin{equation}
\nonumber \includegraphics[scale=0.6,angle=0]{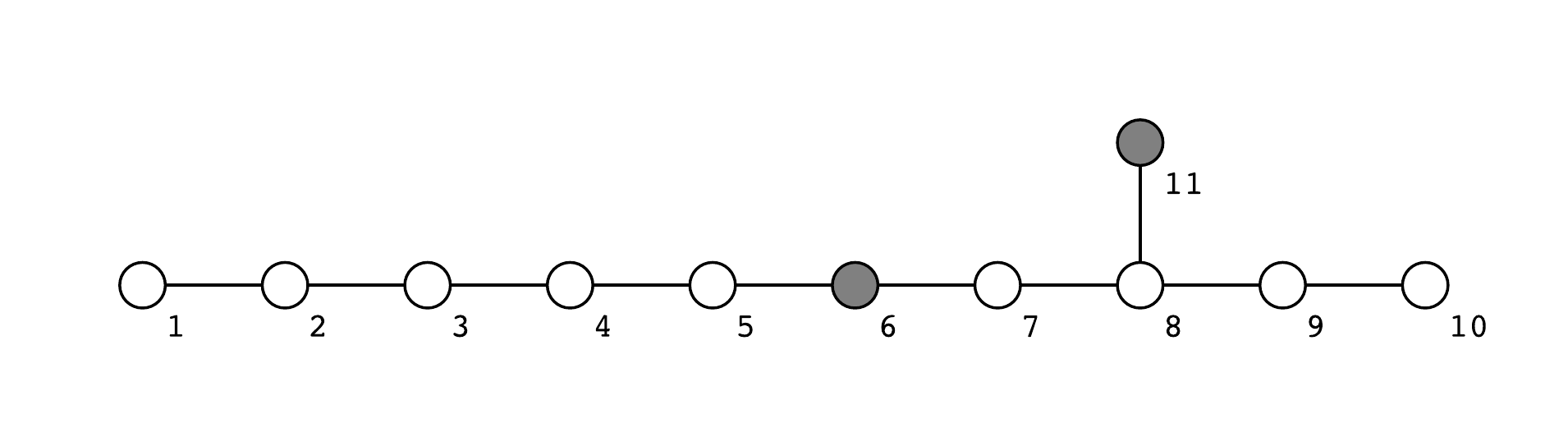}
\end{equation}
This results in the $E_{11}$ generators being decomposed into representations of 
$A_5\otimes A_4\,$, where now the $A_5$ corresponds to the six-dimensional space-time theory, 
while the $A_4$ representations arise from the decomposition of the internal symmetry $SO(10)\,$. 
Once again by dimensionally reducing the low level generators of $E_{11}$ one can identify the set of 
six-dimensional fields which will reduce to the scalar, vector and gravity multiplet of five dimensions. 
The set of generators which give rise to the scalar multiplet ${\bf 42}$ upon dimensional reduction are
\begin{equation}
\hat{\phi}=\{H_{\hat{\mu}} ({\bf 1}), H_{\hat{M}} ({\bf 5}), {K^{\hat{\mu}}}_{\hat{M}} ({\bf 5}),  
{K^{\hat{M}}}_{\hat{N}} ({\bf 10}), R^{\hat{\mu}\hat {M}[2]}  ({\bf 10}), R^{\hat {M}[3]}  ({\bf 10}), 
R^{\hat{\mu}\hat {M}[5]}  ({\bf 1}) \}
\end{equation}
where $\hat{\mu}, \hat{\nu}\in \{1,2,\ldots 6\}$ are the six-dimensional space-time indices 
indicating the $A_5$ tensor structure and $\hat{M}, \hat{N}\in \{1,2,\ldots 5\}$ are the five-dimensional internal indices 
indicating the $A_4$ tensor structure; $H$ are the Cartan sub-algebra elements of $E_{11}$ and we indicate in brackets 
the dimensions of the internal $A_4$ tensor.

Similarly we may identify the six-dimensional origin of the five-dimensional vector and gravity multiplets. 
The vector multiplet which transforms under the ${\bf 27}$ of the internal symmetry is
\begin{equation}
\hat{\phi}^\mu=\{{K^{\hat{\mu}}}_{\hat{\nu}} ({\bf 1}), {K^{\hat{\mu}}}_{\hat{M}} ({\bf 5}),  R^{\hat {\mu}[2]\hat{M}}  
({\bf 5}), R^{\hat{\mu}\hat {M[2]}}  ({\bf 10}), R^{\hat{\mu}[2]\hat {M}[4]}  ({\bf 5}), R^{\hat{\mu}\hat {M}[5]}  
({\bf 1}) \}.
\end{equation}
Nominally the $({\bf 5},{\bf 1})$ arises from the dimensional reduction of
\begin{equation}
\hat{\phi}^{(\mu\nu)}=\{{K^{\hat{\mu}}}_{\hat{\nu}}\}
\end{equation}
although it could equally well arise from the infinite tower of fields which reduce to the five-dimensional dual graviton:
\begin{equation}
\hat{\phi}^{(\mu\nu)}_{\mbox{(m+1)-dual gravity}}=\{{R}^{\hat{\mu}^{(1)}[4] ,\hat{\mu}^{(2)}[4], \ldots, 
\hat{\mu}^{(m)}[4],\hat{\nu}[3],\hat{\rho}}\}.
\end{equation}
The internal indices transform trivially under $A_4$ and have been suppressed. 
These are the set of generators which arise from the dimensional reduction of the eleven dimensional generators of 
$E_{11}$ shown in equation (\ref{11Dmultidualgraviton}). Upon dimensional reduction to five dimensions these generators 
give those indicated in equation (\ref{5Dmultidualgraviton}). 

It has been argued in \cite{Hull:2000zn} that the strong coupling limit of the five-dimensional maximal supergravity 
theory is the superconformal $(4,0)$ theory in six dimensions containing $\bf 27$ self-dual two-forms, 
$\bf 42$ scalars as well as the gravitational degrees of freedom. The scalars reduce trivially to the scalars 
of the five-dimensional theory. The two-form which, as a representation of the little group, in six dimensions carries 
${\bf 6}$ degrees of freedom of which only ${\bf 3}$ are independent due to the self-duality condition in six-dimensions. 
Upon reduction of the two-form one has a choice which degrees of freedom to use to describe the theory: either $\bf 27$ two-forms or $\bf 27$
vectors, which are dual to each other in five-dimensions. 
Instead of the Fierz--Pauli graviton, the gravitational degrees of freedom 
are contained in a mixed symmetry tensor 
$\widehat{C}{}_{\hat{\mu}[2],\hat{\nu}[2]}$ 
which has the symmetries of the Young tableau:
$$\yng(2,2)$$
and, on-shell, has the algebraic properties of the Weyl tensor. 
Because of the self-duality constraint on its curvature, the field carries only five degrees of freedom
and not ten, the dimension of the Weyl tensor representation of the little group in six dimensions. 
Upon dimensional reduction to five dimensions there appear the graviton $h_{\mu\nu}\,$, 
the dual graviton $c_{\mu[2],\nu}$ and the double-dual graviton $d_{\mu[2],\nu[2]}\,$. 
The field $\widehat{C}$ is self-dual in six dimensions, which means that the reduced fields 
are not all independent. Indeed if $\hat{\star}_1 \widehat{C} = \widehat{C}\,$ 
(which implies $\hat{\star}_1\hat{\star}_2 \widehat{C} = \widehat{C}\,$) 
where $\hat{\star}$ is derived from the Hodge dual in six dimensions, 
then upon reduction we have $\star_1 \star_2 d = h$ and 
$\star_2 d = c$ as expected for the dual and double-dual graviton.
Hull referred to this as a triality relation between the three five-dimensional fields $\{h,c,d\}\,$, 
see \cite{Hull:2000zn}.

In terms of $E_{11}$ we understand this as a repackaging of the low-level degrees of 
freedom of the theory into six-dimensional scalars and tensors whose Young tableaux 
have columns of height two, which corresponds to a conformal sector of the 
theory \cite{Siegel:1988gd}. 
This poses a puzzle concerning the mechanism for freezing out the other low level 
generators whose degrees of freedom propagate in the six-dimensional maximal 
supergravity. We will not pursue this here, instead we will comment upon a second problem, namely that 
although the graviton and the dual graviton occur directly 
within the decomposition of $E_{11}\,$, the double-dual graviton field 
$d_{{\mu}[2],{\nu}[2]}\,$, 
transforming trivially under the U-duality group, does not. 
Instead of the double-dual graviton in five-dimensions there is the candidate 
field $\bar{c}_{\mu[3],\nu[2],\rho}$ which may play the same role, and similar 
arguments to those presented in \cite{Hull:2000zn} for the double-dual graviton 
and its lift to six dimensions carry across to the infinite tower of fields 
whose generators are shown in equation (\ref{5Dmultidualgraviton}). 
The fields associated with these generators have Young tableaux:
\begin{equation}
\ytableausetup
{mathmode, boxsize=1.4em}
\begin{ytableau}
\mbox{{\scriptsize n}} & \mbox{{\scriptsize n}} & \none[\dots] & \mbox{{\scriptsize n}}
& \mbox{{\scriptsize n}} &\mbox{{\scriptsize n}} \\
\mbox{{\scriptsize n-1}} & \mbox{{\scriptsize n-1}}  & \none[\dots]& \mbox{{\scriptsize n-1}}  
& \mbox{{\scriptsize n-1}} \\
\none[\vdots] & \none[\vdots]
& \none[\ldots] & \none[\vdots] & \none[\vdots] \\
\scriptstyle 4 & \scriptstyle 4 & \none[\dots]& \scriptstyle 4 
& \scriptstyle 4 \\
\scriptstyle 3 & \scriptstyle 3 & \none[\dots]
& \scriptstyle 3 \\
\end{ytableau}
\end{equation}
where there are $m$ columns of height $(n-2)$ in $n$ dimensions. 
The columns of height $(n-2)$ transform trivially as a representation of the little group in $n$ dimensions. 
In five dimensions when $m=1$ we have the field $\bar{c}_{\mu[3],\nu[2],\rho}$ which carries the same number 
of degrees of freedom as the double-dual graviton. 
Consider a six-dimensional field with the same symmetries 
$\bar{C}_{\hat{\mu}[3],\hat{\nu}[2],\hat{\rho}}\,$. 
We impose that it is self-dual in two independent ways, 
as $\hat{\star}_2 \bar{C} = \bar{C}\,$ and 
$\hat{\star}_1\hat{\star}_3 \bar{C} =\bar{C}\,$, implying 
$\hat{\star}_1\hat{\star}_2\hat{\star}_3 \bar{C}=\bar{C}\,$. Upon reduction to five dimensions we find a large set of fields which are related by the six-dimensional 
dualties including $\bar{c}_{\mu[3],\nu[2],\rho}$ and $c_{\mu[2],\nu}$ together with $e_{\mu[2],\nu[2],\rho}\,$, 
$f_{\mu[3],\nu,\rho}$, $k_{\mu[3],\nu[2]}\,$, $l_{\mu[2],\nu,\rho}\,$, $m_{\mu[2],\nu[2]}$ 
and $n_{\mu[3],\nu}\,$.
Now we note that no graviton appears directly in the reduction but 
$\star_1\star_2 \bar{c}=h$ 
and $\star_1 c =h\,$, so we expect a theory in which there is a choice 
of field used to describe the gravitational 
degrees of freedom. The dualities on the six-dimensional field 
$\bar{C}$ reduce it to carrying $\bf 8$ gravitational 
degrees of freedom of which $\bf 3$ degrees of freedom may be 
eliminated by imposing that the five-dimensional 
field is traceless. 

It is actually possible to find field equations in $6D$ for the 
$\bar{C}_{[3,2,1]}$ field, which yield $\bf 5$ 
propagating degrees of freedom. Starting from the potential 
$\bar{C}_{\hat{\mu}[3],\hat{\nu}[2],\hat{\rho}}\,$, 
one builds the gauge-invariant curvature tensor $K_{[4,3,2]}\,$. 
All the curls of $K_{[4,3,2]}$
vanish, and we propose that the field equations set all the double 
traces of $K$ to zero: ${\rm Tr}^2\,K_{[4,3,2]} = 0\,$, 
with the notation of \cite{Bekaert:2002dt}. 
These kinds of higher-trace field equations were discussed in 
\cite{Hull:2000zn,Hull:2001iu} and an 
analysis of $\mathfrak{gl}_n$-covariant on-shell Hodge duality 
for arbitrary mixed-symmetry gauge fields in Minkoswki space 
can be found in \cite{Bekaert:2002dt}. 
We further impose that some of the single-traces of $K_{[4,3,2]}$
vanish, such that on-shell one has 
${K}_{\hat{\mu}[4],\hat{\nu}[3],\hat{\rho[2]}} =  
\eta_{\hat{\mu}\hat{\rho}}W_{\hat{\mu}[3],\hat{\nu}[3],\hat{\rho}}\,$
where the tensor $W_{[3,3,1]}$ is 
$\mathfrak{so}_6$-irreducible. In other words, it is enough if 
a self-duality condition is imposed on the second column of the
curvature: $*_2 K_{[4,3,2]} = K_{[4,3,2]}\,$. 
By using the general results of \cite{Bekaert:2002dt,Bekaert:2006ix}, 
one can then show that the curvature $W_{[3,3,1]}$ is the gradient 
of the curvature $\widehat{K}_{[3,3]}$ for the self-dual gauge field 
$\widehat{C}{}_{[2,2]}$ discussed by Hull \cite{Hull:2000zn}
and introduced in \cite{Strathdee:1987p5523}. 
Note that, by using the equations
${K}_{\hat{\mu}[4],\hat{\nu}[3],\hat{\rho[2]}} = 
\eta_{\hat{\mu}\hat{\rho}}W_{\hat{\mu}[3],\hat{\nu}[3],\hat{\rho}}\,$ 
for a traceless $W_{[3,3,1]}\,$, 
the relation $\hat{\star}_1\hat{\star}_3 \bar{C} =\bar{C}$
proposed above is indeed satisfied. 
With the above field equations, the $\bar{C}_{[3,2,1]}$ field 
in six-dimension propagates the same $\bf 5$ degrees
of freedom as does $\widehat{C}{}_{[2,2]}\,$, 
which is what we wanted to show.  

We conclude this section with some comments on the construction of Young tableaux of $E_{11}$ generators, which will be 
analogous to the auxiliary $Z$ fields appearing in the earlier sections of this note, required for a covariant formulation of the 
double-dual graviton. Generalised Kac-Moody algebras are constructed from their Cartan matrix $A$ together with the Serre 
relations:
\begin{equation}
\overbrace{[E_a,[E_a,\ldots [E_a}^{1-A_{ab}}, E_b] \ldots]]=0 \qquad \mbox{and} \qquad
\underbrace{[F_a,[F_a,\ldots [F_a}_{1-A_{ab}}, F_b] \ldots]]=0
\label{Serre}
\end{equation}
where $E_a$($F_a$) are the positive(negative) generators of the algebra, $A_{ab}$ is a Cartan matrix entry and there are 
$(1-A_{ab})$  $E_a$ or $F_a$ generators in each relation. These relations give constraints\footnote{Roots of $E_{11}$ have 
length squared which is bounded from above and if this is normalised to two then $\beta^2=2,0,-2,-4,\ldots$.} on the root 
length of the roots associated with the algebra which may be directly related to the Young tableaux of the generators of the 
algebra \cite{Cook:2009ri}. For $E_{11}$ a generic root $\vec{\beta}=\sum_{i=1}^{11} w_i \vec{e}_i$ 
 is associated with a 
generator whose Young tableau has rows of width $w_i$. The root length squared of this root is
\begin{equation}
\beta^2 =\sum_{i=1}^{11} ({w}_i)^2-L^2 
\end{equation}
where $L\equiv \frac{1}{3}\sum_{i=1}^{11}w_i$ is the level the generator appears at in the decomposition of $E_{11}$ and is one-third of the number of boxes in its Young tableau. The root length formula is such that if one moves a single box of a mixed-symmetry Young tableau one column to the left the root length squared is reduced by two. Suppose that one moves a single box in this way, this corresponds to $w_k\rightarrow w_k-1$ and $w_l\rightarrow w_l+1$ (for some row k and another row l) and $L\rightarrow L$. The root length squared changes as
\begin{equation}
\beta^2 \rightarrow \beta'^2=\sum_{i=1}^{11} ({w}_i)^2-2w_k+2w_l+2-L^2 =\beta^2 -2
\end{equation}
where we have used the observation that $w_k=w_l+2$ since the box is moved from one column to the top of the adjacent column to the left. This is a useful observation as given a real root associated with a mixed symmetry Young tableau one can identify a sequence of null and imaginary roots in the algebra whose generators have the symmetries of Young tableaux formed by repeatedly moving boxes to the left. For example, the five-dimensional field $\bar{c}_{\mu[3]\nu[2]\rho}$ is associated with a real root appearing at level six in the decomposition of $E_{11}$, while at the same level there appear null and imaginary roots also transforming trivially under the internal $E_6$ symmetry, which are derived by moving boxes to the left in the Young tableau as shown in table \ref{TableofE11doubledualgenerators}. 
\begin{table}[h]
\centering 
\scalebox {0.95} {
\begin{tabular}{c | c | c | c } 
Root length & Field & Multiplicity & Outer\\
squared, $\beta^2$ & &  & multiplicity, $\mu$ \\
 \hline  &&&\\
2 &$\tiny\yng(3,2,1)$ &1&1\\
 &$\bar{c}_{\mu[3],\nu[2],\rho}$ & &\\
 \hline &&&\\
0&$\tiny\yng(3,1,1,1)$&8&1 \\
 &${q_{\mu[4],\nu,\rho}}$ &&\\
  \hline &&&\\
0& $\tiny\yng(2,2,2)$ &8&1 \\
 &$r_{\mu[3],\nu[3]}$ &&\\
 \hline &&&\\
-2&$\tiny\yng(2,2,1,1)$ &44&4 \\
 &$s_{\mu[4],\nu[2]}$&&\\
 \hline &&&\\
-4&$\tiny\yng(2,1,1,1,1)$  &206 & 5\\
 &$t_{\mu[5]\nu}$ && 
\end{tabular} }
\caption{The fields of $E_{11}$ reduced to five dimensions with six tensor indices which transform trivially under the internal $E_6$ symmetry.}\label{TableofE11doubledualgenerators}
\end{table}

This is the same pattern that occurred in the gauge fields used to 
construct the action for the dual graviton tower of gravitational degrees of freedom highlighted in section 3. The decomposition of the first of these fields is shown in equation (\ref{C1}), where the off-shell gauge fields required to guarantee that the Curtright field propagates the correct number of gravitational degrees of freedom are shown. It is not a coincidence that the sets of fields identified within $E_{11}$ and those required to constrain the off-shell degrees of freedom are the same. In both cases it is the same consideration of identifying the irreducible highest weight representations that singles out the sets of fields. So we expect the gauge fields required to construct the Curtright action for the tower of 
dual graviton
fields  to match the fields which appear in $E_{11}$ at the same level as the multi-dual graviton generator.
This leads to the tantalising possibility that null and imaginary roots 
of Kac-Moody algebras may be associated with gauge fields, and gauge for gauge fields and so on.

\section{Conclusion}
\label{sec:Conclusion}
In this paper we constructed an explicit linearised action for the 
double-dual graviton given in equations (\ref{divergenceofD}-\ref{acoeff}) for arbitrary dimension, 
$n\geqslant 5\,$. The off-shell Hodge dualisation that transforms the linearised graviton $h_{\mu\nu}$ into the dual graviton $C_{\mu[n-3],\nu}$ and the double-dual graviton $D_{\mu[n-3],\nu[n-3]}$ were understood at the level of their corresponding actions. This was done by first constructing a parent action $S[\Omega,Y]$ where the algebraic elimination of one set of fields reduced the parent action to the Curtright action but by eliminating the other set of fields the action for the double-dual graviton remained. This procedure had previously been used to construct a parent action which related the Fierz--Pauli action to the Curtright action \cite{Boulanger:2003vs}. We indicate the series of parent actions and their algebraic reductions by the sketch below.
\begin{align*}
&\qquad S[\Omega_{a[2],b},Y_{a[3],b}] \qquad \quad S[H_{a[n-3],b[2]},D_{b[3],a[n-3]}]\\
&\qquad \swarrow \qquad\qquad  \searrow \quad \quad  \quad \swarrow \qquad\qquad  \searrow \\
&S_{\rm FP} (h_{\mu\nu})  \qquad S_{\rm Curt.}(C_{\mu[n-3],\nu})  \qquad S_{\rm DD}(D_{\mu[n-3],\nu[n-3]})
\end{align*}
While the off-shell dualisation trivially transforms the fundamental 
gravity field the transformation of the actions is not so simple and an 
increasingly complicated set of auxiliary fields are required to 
construct the action for each subsequent dualisation. The set of fields 
required to guarantee that the double-dual graviton propagates the 
correct number of degrees of freedom is given in equation 
(\ref{spectrum}). There are two novel aspects that appear in the 
construction of the action for the double-dual graviton.
\begin{itemize}
\item[(i.)] 
The double-dual action retains a number of mixed-symmetry 
fields on top of the $\mathfrak{gl}_n$-irreducible $[n-3,n-3]$ field. 
These extra fields are required for consistency of the action 
principle and correct number of propagating degrees of freedom.
They are not auxiliary in the sense that their equations of motion
cannot be solved algebraically in terms of the $[n-3,n-3]$ field, 
however they can be eliminated on-shell by differential gauge 
symmetries. The detailed mechanism will be explained elsewhere 
\cite{BP} using the frame formulation. 
Such a situation did not occur for the parent actions of 
\cite{Boulanger:2003vs}. 
\item[(ii.)]  
The action retains an unfixed parameter $\beta\,$, 
which corresponds to the possibility of adding total derivative terms 
to the Curtright Lagrangian when expressed in terms of the 
curl $H_{[n-3,2]}(C)\,$. The relation between two different
child actions can be viewed as a Legendre transformation.
At certain values of $\beta$, the Legendre transformation becomes 
non-invertible. This is what happens for the two values 
$\beta^{(1)}$ and $\beta^{(2)}$ given after (\ref{acoeff}).
Fixing a non-singular value for $\beta$ in (\ref{acren}) 
will correspondingly fix it in the Lagrangian of (\ref{fin2}). 
\end{itemize}

In addition to the double-dual graviton there exists an infinite tower 
of dual fields each carrying the gravitational degrees of freedom, this 
was observed in section \ref{sec:Mtimes}. These infinite towers carry
representations of the little group $SO(n-2)$ in $n$ dimensions
which are non-trivial and all equivalent to the spin-2 $SO(n-2)$-irrep 
through multiplication/contraction by a number of $SO(n-2)$ metric 
tensors and antisymmetric $SO(n-2)$ symbols. Correspondingly, the 
$\mathfrak{gl}_n$-covariant field equations will feature high powers of 
the trace operations on the curvature tensor $K\,$, as was proposed 
in \cite{Hull:2001iu,Bekaert:2002dt}.\footnote{The double-dual graviton
introduced by Hull \cite{Hull:2000rr,Hull:2001iu} is the prototype
for such a situation and partly motivated the general analysis 
presented in \cite{Bekaert:2002dt}.}
In this context, the recent paper \cite{Henneaux:2011mm} also 
considered the $\mathfrak{g}^{+++}$ tower of fields obtained by 
attaching columns of length $(n-2)$ 
to the Young diagrams associated with 
a finite-dimensional algebra $\mathfrak{g}\,$, and 
under some assumptions concluded that 
these fields carry no propagating degrees of freedom.
We think that the present analysis might be relevant 
in this context, as we have seen, for example, that the double-dual graviton 
possesses, as a subset of its complete set
of gauge parameters, those of a type-$[n-3,n-3]$ 
Labastida-like gauge field. That covariant equations of motion
relevant for mixed--symmetry gauge fields can 
bring in higher powers of the traces of the 
curvature tensors\footnote{We note that the formulation 
${\rm Tr} \,K=0$ of the Labastida equations was found in 
\cite{Bekaert:2003az}. It is thanks to this reformulation 
of the Labastida equations that it became possible to prove 
that the Labastida equations are actually correct, 
ensuring unitarity and absence of ghosts,   
see \cite{Bekaert:2006ix} and detailed discussions therein.}
is a feature that might be taken into account
in trying to build an $E_{11}$-invariant off-shell 
formulation of supergravity. 
In this work we did not investigate the possibility 
of building an action invariant under $E_{11}\,$ and such that
only one graviton would propagate on-shell in the spin-2 sector,  
but we hope that our results can be useful for that goal. 
In any case, our results show that the Labastida 
formulation is not the only relevant one for such a purpose.

We characterise three towers of fields as the Fierz--Pauli tower shown in equation (\ref{FPtower}) which is derived from the linearised graviton, the dual graviton tower shown in equation (\ref{Curttower}) which is derived from the dual graviton and the double-dual tower shown in equation (\ref{Doubledualtower}) which is derived from the double-dual graviton. The dual graviton tower was first recognised as an inifinite set of dual graviton fields in \cite{Riccioni:2006az} where they were identified within the algebra of $E_{11}$. In this paper we presented the steps required to construct the linearised action associated with any of the fields in each of the three gravity towers. The set of fields required to write down the off-shell action grows with the number of dualisations needed to relate the field to the graviton. The actions, although they propagate the same degrees of freedom as the graviton, contain many more fields. In \cite{Riccioni:2006az} towers of fields dual to the membrane and the fivebrane gauge fields of eleven-dimensional supergravity were also found within $E_{11}$ and dual actions can be constructed in the same manner for these infinite sets of fields as for the dual graviton tower using the prescription given in this paper.

The work presented here was, in part, inspired by the argument that the strong coupling limit of five dimensional maximal supergravity contains the double-dual graviton \cite{Hull:2000zn}. It was expected that the corner of M-theory identified by the strong coupling limit with a six-dimensional $(4,0)$ superconformal theory would be discovered within $E_{11}$. 
However 
no such double-dual graviton is contained within $E_{11}$, instead the dual graviton tower of multi-dual graviton fields \cite{Riccioni:2006az} are the only candidate dual gravity fields. We showed in section \ref{sec:E11 notes} that first of the dual gravity fields, $\bar{c}_{\mu[3]\nu[2]\rho}$ in five dimensions, whose symmetries are described by the Young tableau $\tiny\yng(3,2,1)$ propagate the same $\bf{5}$ degrees of freedom as the double-dual graviton $\bar{d}_{\mu[2]\nu[2]}$ whose symmetries are those of the Young tableau $\tiny\yng(2,2)$. Indeed $\bar{c}_{\mu[3]\nu[2]\rho}$ is singled out as it transforms trivially under the internal $E_6$ symmetry (as seen from the decomposition of $E_{11}$ to five dimensions). 
We noted on-shell that the gauge field might have a number of dualities imposed upon it 
namely $\bar{C}_{[3,2,1]}$ satisfy $\hat{\star}_2 \bar{C} = \bar{C}\,$, 
$\hat{\star}_1\hat{\star}_3 \bar{C} =\bar{C}$ and 
$\hat{\star}_1\hat{\star}_2\hat{\star}_3 \bar{C}=\bar{C}\,$.
The requirement that the following six-dimensional field 
equations are imposed
\begin{equation}
{\rm Tr}^2\,K_{[4,3,2]} = 0\ \;, \quad {K}_{\hat{\mu}[4],\hat{\nu}[3],\hat{\rho[2]}} =  
\eta_{\hat{\mu}\hat{\rho}}W_{\hat{\mu}[3],\hat{\nu}[3],\hat{\rho}} 
\quad {\rm and} \quad 
K_{[4,3,2]} = *_2 K_{[4,3,2]} \quad, 
\end{equation}
where $K_{[4,3,2]}$ is the field strength 
\cite{Hull:2001iu,Bekaert:2002dt} of $\bar{C}_{[3,2,1]}\,$, 
is sufficient to ensure that the strong coupling lift of 
$\bar{c}_{[3,2,1]}$ to $\bar{C}_{[3,2,1]}$ in six dimensions 
preserves the $E_6$ multiplets as well as the gravitational 
degrees of freedom. 
These are similar to the arguments which were made by Hull 
\cite{Hull:2000zn} for the double-dual graviton.

It was emphasised above that the linearised action for the double-dual graviton retained a number of supplementary mixed-symmetry fields. These mixed symmetry fields were identified in the construction of the action from the irreducible components of double-dual graviton which could not be completely gauged away by algebraic gauge transformations. The same is true of the first field in the dual graviton tower of dualities, namely in five-dimensions the $\tiny\yng(3,2,1)$ and a set of mixed symmetry fields expected to be retained in the corresponding action are listed in equation (\ref{C1}). The construction of $E_{11}$ rests upon the Serre relations, see equation (\ref{Serre}), which guarantee the irreducibility of any representation of $E_{11}$. The irreducibility restricts which mixed-symmetry tensors appear in the decomposition of $E_{11}$ to tensor representations of $\mathfrak{sl}_{11-n}$ relevant to $n$ dimensional extended supergravity. It is not coincidental that the same sets of mixed symmetry fields required for the action of the $\tiny\yng(3,2,1)$ in five dimensions are also contained within $E_{11}$. It is the same consideration of irreducibility that has been used both in the programme for constructing dual actions and in the definition of $E_{11}$. The corresponding supplementary
mixed-symmetry tensors are associated with null and imaginary roots in the root system of $E_{11}$.

\section*{Acknowledgements}
\label{sec:Acknowledgements}

N.B. acknowledges X. Bekaert, S. Cnockaert, M. Henneaux and O. Hohm for 
early discussions, and thanks King's College London and the 
Erwin Schr\"odinger Institute in Vienna for kind hospitality. 
He also gratefully thanks A. Kleinschmidt for discussions related to 
the paper \cite{Englert:2007qb}
and M. Henneaux for his comments and explanations related to the
work \cite{Henneaux:2011mm}.
N.B. and D.P. thank E. D. Skvortsov, Ph. Spindel and P. Sundell for their comments. 
N.B., D.P. and P.P.C. thank P. West for his comments on a final draft of the paper. 
P.P.C. is grateful to the organisers of the {\it Mathematics and Applications of String and M-theory} 
workshop at the Isaac Newton Institute Cambridge where part of this work was carried out. 
The work of N.B. and D.P. was supported in part by an ARC contract No. AUWB-2010-10/15-UMONS-1. 
P.P.C. is supported by the S.T.F.C. rolling grant (ST/G00395/1) of the theoretical physics group 
at King's College London.


\begin{thebibliography}{10}

\bibitem{West:2001as}
P.~C. West, ``{E(11) and M theory},''
  \href{http://dx.doi.org/10.1088/0264-9381/18/21/305}{{\em Class.Quant.Grav.}
  {\bf 18} (2001)  4443--4460},
\href{http://arxiv.org/abs/hep-th/0104081}{{\tt arXiv:hep-th/0104081
  [hep-th]}}.

\bibitem{Damour:2000hv}
T.~Damour and M.~Henneaux, ``{E(10), BE(10) and arithmetical chaos in
  superstring cosmology},''
  \href{http://dx.doi.org/10.1103/PhysRevLett.86.4749}{{\em Phys.Rev.Lett.}
  {\bf 86} (2001)  4749--4752},
\href{http://arxiv.org/abs/hep-th/0012172}{{\tt arXiv:hep-th/0012172
  [hep-th]}}.

\bibitem{Damour:2002et}
T.~Damour, M.~Henneaux, and H.~Nicolai, ``{Cosmological billiards},'' {\em
  Class.Quant.Grav.} {\bf 20} (2003)  R145--R200,
\href{http://arxiv.org/abs/hep-th/0212256}{{\tt arXiv:hep-th/0212256
  [hep-th]}}.

\bibitem{Julia:1980gr}
B.~Julia, ``Group disintegrations,''
{\em Conf.Proc.} {\bf C8006162} (1980)  331--350.

\bibitem{Julia:1997cy}
B.~Julia, ``{Dualities in the classical supergravity limits: Dualizations,
  dualities and a detour via (4k+2)-dimensions},''
  \href{http://arxiv.org/abs/hep-th/9805083}{{\tt arXiv:hep-th/9805083
  [hep-th]}}.
latex 22p. to appear in Proceedings of Cargese NATO ASI: Strings, branes and
  dualities May 1997 ed. P. Windey Report-no: LPTENS-98/07.

\bibitem{Boulanger:2003vs}
N.~Boulanger, S.~Cnockaert, and M.~Henneaux, ``{A note on spin s duality},''
  {\em JHEP} {\bf 0306} (2003)  060,
\href{http://arxiv.org/abs/hep-th/0306023}{{\tt arXiv:hep-th/0306023
  [hep-th]}}.

\bibitem{Curtright:1980yk}
T.~Curtright, ``Generalized gauge fields,''
  \href{http://dx.doi.org/10.1016/0370-2693(85)91235-3}{{\em Phys.Lett.} {\bf
  B165} (1985)  304}.

\bibitem{Aulakh:1986cb}
C.~Aulakh, I.~Koh, and S.~Ouvry, ``Higher spin fields with mixed symmetry,''
  \href{http://dx.doi.org/10.1016/0370-2693(86)90518-6}{{\em Phys.Lett.} {\bf
  B173} (1986)  284}.

\bibitem{Fierz:1939ix}
M.~Fierz and W.~Pauli, ``{On relativistic wave equations for particles of
  arbitrary spin in an electromagnetic field},''
{\em Proc.Roy.Soc.Lond.} {\bf A173} (1939)  211--232.

\bibitem{Boulanger:2008nd}
N.~Boulanger and O.~Hohm, ``{Non-linear parent action and dual gravity},''
  \href{http://dx.doi.org/10.1103/PhysRevD.78.064027}{{\em Phys.Rev.} {\bf D78}
  (2008)  064027},
\href{http://arxiv.org/abs/0806.2775}{{\tt arXiv:0806.2775 [hep-th]}}.

\bibitem{West:2002jj}
P.~C. West, ``{Very extended E(8) and A(8) at low levels, gravity and
  supergravity},'' \href{http://dx.doi.org/10.1088/0264-9381/20/11/328}{{\em
  Class.Quant.Grav.} {\bf 20} (2003)  2393--2406},
\href{http://arxiv.org/abs/hep-th/0212291}{{\tt arXiv:hep-th/0212291
  [hep-th]}}.

\bibitem{Hull:2000rr}
C.~Hull, ``{Symmetries and compactifications of (4,0) conformal gravity},''
  {\em JHEP} {\bf 0012} (2000)  007,
\href{http://arxiv.org/abs/hep-th/0011215}{{\tt arXiv:hep-th/0011215
  [hep-th]}}.

\bibitem{Hull:2000zn}
C.~Hull, ``{Strongly coupled gravity and duality},''
  \href{http://dx.doi.org/10.1016/S0550-3213(00)00323-0}{{\em Nucl.Phys.} {\bf
  B583} (2000)  237--259},
\href{http://arxiv.org/abs/hep-th/0004195}{{\tt arXiv:hep-th/0004195
  [hep-th]}}.

\bibitem{Hull:2001iu}
C.~Hull, ``{Duality in gravity and higher spin gauge fields},'' {\em JHEP} {\bf
  0109} (2001)  027,
\href{http://arxiv.org/abs/hep-th/0107149}{{\tt arXiv:hep-th/0107149
  [hep-th]}}.

\bibitem{Riccioni:2006az}
F.~Riccioni and P.~C. West, ``{Dual fields and E(11)},''
  \href{http://dx.doi.org/10.1016/j.physletb.2006.12.050}{{\em Phys.Lett.} {\bf
  B645} (2007)  286--292},
\href{http://arxiv.org/abs/hep-th/0612001}{{\tt arXiv:hep-th/0612001
  [hep-th]}}.

\bibitem{Chiodaroli:2011pp}
M.~Chiodaroli, M.~Gunaydin, and R.~Roiban, ``{Superconformal symmetry and
  maximal supergravity in various dimensions},'' {\em JHEP} {\bf 1203} (2012)
  093, \href{http://arxiv.org/abs/1108.3085}{{\tt arXiv:1108.3085 [hep-th]}}.

\bibitem{Englert:2007qb}
F.~Englert, L.~Houart, A.~Kleinschmidt, H.~Nicolai, and N.~Tabti, ``{An E(9)
  multiplet of BPS states},''
  \href{http://dx.doi.org/10.1088/1126-6708/2007/05/065}{{\em JHEP} {\bf 0705}
  (2007)  065},
\href{http://arxiv.org/abs/hep-th/0703285}{{\tt arXiv:hep-th/0703285
  [hep-th]}}.

\bibitem{Geroch:1970nt}
R.~P. Geroch, ``{A Method for generating solutions of Einstein's equations},''
\href{http://dx.doi.org/10.1063/1.1665681}{{\em J.Math.Phys.} {\bf 12} (1971)
  918--924}.

\bibitem{Geroch:1972yt}
R.~P. Geroch, ``{A Method for generating new solutions of Einstein's equation.
  2},''
\href{http://dx.doi.org/10.1063/1.1665990}{{\em J.Math.Phys.} {\bf 13} (1972)
  394--404}.

\bibitem{Breitenlohner:1986um}
P.~Breitenlohner and D.~Maison, ``{On the Geroch Group},''
{\em Annales Poincare Phys.Theor.} {\bf 46} (1987)  215.

\bibitem{Damour:2002cu}
T.~Damour, M.~Henneaux, and H.~Nicolai, ``{E(10) and a 'small tension
  expansion' of M theory},''
  \href{http://dx.doi.org/10.1103/PhysRevLett.89.221601}{{\em Phys.Rev.Lett.}
  {\bf 89} (2002)  221601},
\href{http://arxiv.org/abs/hep-th/0207267}{{\tt arXiv:hep-th/0207267
  [hep-th]}}.

\bibitem{Bekaert:2002dt}
X.~Bekaert and N.~Boulanger, ``{Tensor gauge fields in arbitrary
  representations of GL(D,R): Duality and Poincare lemma},''
  \href{http://dx.doi.org/10.1007/s00220-003-0995-1}{{\em Commun. Math. Phys.}
  {\bf 245} (2004)  27--67},
\href{http://arxiv.org/abs/hep-th/0208058}{{\tt arXiv:hep-th/0208058}}.

\bibitem{Zinoviev:2003ix}
Y.~M. Zinoviev, ``{First order formalism for mixed symmetry tensor fields},''
  \href{http://arxiv.org/abs/hep-th/0304067}{{\tt arXiv:hep-th/0304067
  [hep-th]}}.

\bibitem{Skvortsov:2008vs}
E.~D. Skvortsov, ``{Mixed-Symmetry Massless Fields in Minkowski space
  Unfolded},'' \href{http://dx.doi.org/10.1088/1126-6708/2008/07/004}{{\em
  JHEP} {\bf 07} (2008)  004},
\href{http://arxiv.org/abs/0801.2268}{{\tt arXiv:0801.2268 [hep-th]}}.

\bibitem{Skvortsov:2008sh}
E.~Skvortsov, ``{Frame-like Actions for Massless Mixed-Symmetry Fields in Minkowski
	space},'' 
    \href{http://dx.doi.org/10.1016/j.nuclphysb.2008.09.007}{{\em Nucl.Phys.}
  {\bf B808} (2009) 569-591}, \href{http://arxiv.org/abs/0807.0903}{{\tt
  arXiv:0807.0903 [hep-th]}}.

\bibitem{Skvortsov:2010nh}
E.~Skvortsov and Y.~Zinoviev, ``{Frame-like Actions for Massless Mixed-Symmetry
  Fields in Minkowski space. Fermions},''
  \href{http://dx.doi.org/10.1016/j.nuclphysb.2010.10.012}{{\em Nucl.Phys.}
  {\bf B843} (2011)  559--569}, \href{http://arxiv.org/abs/1007.4944}{{\tt
  arXiv:1007.4944 [hep-th]}}.

\bibitem{Fradkin:1984ai}
E.~Fradkin and A.~A. Tseytlin, ``Quantum equivalence of dual field theories,''
\href{http://dx.doi.org/10.1016/0003-4916(85)90225-8}{{\em Annals Phys.} {\bf
  162} (1985)  31}.

\bibitem{Weyl:1929fm}
H.~Weyl, ``{Electron and Gravitation. 1. (In German)},''
{\em Z.Phys.} {\bf 56} (1929)  330--352.

\bibitem{Bekaert:2002uh}
X.~Bekaert, N.~Boulanger, and M.~Henneaux, ``{Consistent deformations of dual
  formulations of linearized gravity: A No go result},''
  \href{http://dx.doi.org/10.1103/PhysRevD.67.044010}{{\em Phys.Rev.} {\bf D67}
  (2003)  044010}, \href{http://arxiv.org/abs/hep-th/0210278}{{\tt
  arXiv:hep-th/0210278 [hep-th]}}.

\bibitem{Bekaert:2004dz}
X.~Bekaert, N.~Boulanger, and S.~Cnockaert, ``{No self-interaction for
  two-column massless fields},''
  \href{http://dx.doi.org/10.1063/1.1823032}{{\em J.Math.Phys.} {\bf 46} (2005)
   012303}, \href{http://arxiv.org/abs/hep-th/0407102}{{\tt
  arXiv:hep-th/0407102 [hep-th]}}.

\bibitem{Riccioni:2007au}
F.~Riccioni and P.~C. West, ``{The E(11) origin of all maximal
  supergravities},''
  \href{http://dx.doi.org/10.1088/1126-6708/2007/07/063}{{\em JHEP} {\bf 0707}
  (2007)  063},
\href{http://arxiv.org/abs/0705.0752}{{\tt arXiv:0705.0752 [hep-th]}}.

\bibitem{Riccioni:2009xr}
F.~Riccioni, D.~Steele, and P.~West, ``{The E(11) origin of all maximal
  supergravities: The Hierarchy of field-strengths},''
  \href{http://dx.doi.org/10.1088/1126-6708/2009/09/095}{{\em JHEP} {\bf 0909}
  (2009)  095},
\href{http://arxiv.org/abs/0906.1177}{{\tt arXiv:0906.1177 [hep-th]}}.

\bibitem{Riccioni:2008jz}
F.~Riccioni, A.~Van~Proeyen, and P.~C. West, ``{Real forms of very extended
  Kac-Moody algebras and theories with eight supersymmetries},''
  \href{http://dx.doi.org/10.1088/1126-6708/2008/05/079}{{\em JHEP} {\bf 0805}
  (2008)  079},
\href{http://arxiv.org/abs/0801.2763}{{\tt arXiv:0801.2763 [hep-th]}}.

\bibitem{Strathdee:1987p5523}
J.~Strathdee, ``Extended poincare supersymmetry,'' {\em Int. J. Mod. Phys. A}
  {\bf 2} (1987) no.~1, 273.

\bibitem{Englert:2003p595}
F.~Englert, L.~Houart, A.~Taormina, and P.~West, ``The symmetry of
  m-theories,''{\em Journal of High Energy Physics} {\bf 0309} (Sep, 2003)  ,
  \href{http://arxiv.org/abs/hep-th/0304206v2}{{\tt hep-th/0304206v2}}.
  \url{http://arxiv.org/abs/hep-th/0304206v2}.

\bibitem{Siegel:1988gd}
W.~Siegel, ``All free conformal representations in all dimensions,''
\href{http://dx.doi.org/10.1142/S0217751X89000819}{{\em Int.J.Mod.Phys.} {\bf
  A4} (1989)  2015}.

\bibitem{Bekaert:2006ix}
X.~Bekaert and N.~Boulanger, ``{Tensor gauge fields in arbitrary
  representations of GL(D,R). II: Quadratic actions},''
  \href{http://dx.doi.org/10.1007/s00220-006-0187-x}{{\em Commun. Math. Phys.}
  {\bf 271} (2007)  723--773},
\href{http://arxiv.org/abs/hep-th/0606198}{{\tt arXiv:hep-th/0606198}}.

\bibitem{Cook:2009ri}
P.~P. Cook, ``{Exotic E(11) branes as composite gravitational solutions},''
  \href{http://dx.doi.org/10.1088/0264-9381/26/23/235023}{{\em
  Class.Quant.Grav.} {\bf 26} (2009)  235023},
\href{http://arxiv.org/abs/0908.0485}{{\tt arXiv:0908.0485 [hep-th]}}.

\bibitem{BP}
N.~Boulanger and D.~Ponomarev. In preparation.

\bibitem{Henneaux:2011mm}
M.~Henneaux, A.~Kleinschmidt, and H.~Nicolai, ``{Higher spin gauge fields and
  extended Kac-Moody symmetries},''
\href{http://arxiv.org/abs/1110.4460}{{\tt arXiv:1110.4460 [hep-th]}}.

\bibitem{Bekaert:2003az}
X.~Bekaert and N.~Boulanger, ``{On geometric equations and duality for free
  higher spins},'' \href{http://dx.doi.org/10.1016/S0370-2693(03)00409-X}{{\em
  Phys. Lett.} {\bf B561} (2003)  183--190},
\href{http://arxiv.org/abs/hep-th/0301243}{{\tt arXiv:hep-th/0301243}}.

\end{thebibliography}

\providecommand{\href}[2]{#2}\begingroup\raggedright\endgroup

\end{document}